\documentstyle[aps,preprint,tighten]{revtex}
\begin {document}
\draft

\title{ Precise solution of few-body problems
with stochastic variational method
on correlated Gaussian basis}

\author{K. Varga$^{1,2,3}$ and Y. Suzuki$^{1}$
\
\
\\$^{1}$ Department of Physics, Niigata University, Niigata 950-21, Japan
\\$^{2}$ RIKEN, Hirosawa, Wako, Saitama 351-01, Japan
\\and
\\ $^{3}$Institute of Nuclear Research of the Hungarian Academy of Sciences,
\\
Debrecen, H--4001, Hungary}
\date{\today}
\maketitle
\begin{abstract}
Precise variational solutions are given for problems involving diverse
fermionic and bosonic $N=2-7$-body systems. The trial wave functions
are chosen to be combinations of correlated Gaussians, which are
constructed from products of the single-particle Gaussian wave packets
through an integral transformation, thereby facilitating fully analytical
calculations of the matrix elements.  The nonlinear parameters of the trial
function are chosen by a stochastic technique.
  The method has proved very efficient, virtually exact, and it seems
feasible for any few-body bound-state problems emerging in nuclear or atomic
physics.
\end{abstract}
\vskip 1.cm
\narrowtext
\section{Introduction}
Few-body problems of interacting particles have vital importance in all
branches of physics from hadron to celestial levels. The main interest
in the few-body problems lies in, e.g., finding an accurate solution for the
system,
testing the equation of motion and the conservation laws and
symmetries, or looking for unknown interactions governing the system.
\par\indent
   The investigation of few-nucleon systems interacting via realistic
forces has always been in the center of the interest. Considerable effort
has been exerted to obtain
accurate ground-state properties of the few-nucleon systems with
Faddeev-Yakubovsky (FY) \cite{payne,faddeev,glockb}, variational
\cite{atms,hypers,crcbgv},  variational Monte Carlo (VMC)
\cite{mcv,vmc} and Green's function Monte
Carlo (GFMC) methods \cite{gfma,gfmb}. Most of these approaches has focused
on three- or four-body problems.
\par\indent
To treat an $N$-particle system, one needs to cope with a large number of
variables
required to specify the wave function. By using $(N\!\!-\!\!1)$ relative
coordinates to
describe the system, for example, the discretization on a mesh with
$p$ points, or the expansion of the function of the relative motion between
the particles in terms
of $p$ suitably chosen functions leads to $p^{(N-1)}$ mesh points or basis
functions, which becomes prohibitively large with increasing $N$.
All but the Monte Carlo methods face this difficulty
as the number of particles increases. The VMC and GFMC
methods have proved to be most successful by being able to go beyond
the four-nucleon problem \cite{prl95}.  The secret of the efficiency of the
Monte Carlo methods is the use of an importance sampling
 of the most relevant parts of the configuration space.
This fact naturally raises a question: Even if the wave function of the
$N$-particle system is expanded into an (excessively) large number of basis
functions, can't one reduce the problem to a tractable one by selecting
``the most important'' basis functions ?
\par\indent
The aim of this paper is to present an alternative variational approach, the
 stochastic variational method (SVM) \cite{SVM,VSL}, by using the correlated
Gaussians as basis functions \cite{corrgauss,temper}. Examples whose
solutions were known before are used to demonstrate the
performance of the method in treating nuclear as well as Coulomb
interactions.
To highlight some new physics, we have also included problems that have
been hitherto unsolved.  We give the formulation and some details of the
method of the calculation and show applications to $N=2-7$-particle systems.
\par\indent
   The variational foundation for the time-independent Schr\"odinger equation
provides a solid and arbitrarily improvable framework for the solution of
bound-state problems. The crucial point of the variational approach is the
choice of the trial function. There are two widely applied strategies:
(1) to select the most appropriate functional form to describe the
short-range as well as long-range correlations and to compute the matrix
elements by Monte Carlo technique, or (2) to use a number, possibly a great
number, of simple terms, which facilitate the analytical calculation of the
matrix elements. We
follow the second course by using an expansion over a correlated Gaussian
``basis".
\par\indent
To solve the $N$-particle problem, it is of prime importance to describe the
correlation between the
particles properly. The correlation is conveniently represented by a
correlation factor, $F=\prod_{i<j}^{N}f_{ij}$
\cite{atms,hypers,mcv,vmc,etbm}.
Most calculations have used this form of $F$ directly to evaluate
the matrix elements. Such calculations are, however, fairly
involved beyond the three-particle system and performed by
Monte Carlo integrations. An alternative way to incorporate the
correlation is to approximate $f_{ij}$ as a linear combination of Gaussians
${\rm exp}(-\alpha_{ij} ({\bf r}_i-{\bf r}_j)^2)$.
The $N$-particle basis function then contains product of these
Gaussians:
$\prod_{i<j}^{N}{\rm exp}(-\alpha_{ij}({\bf r}_i-{\bf r}_j)^2)=
{\rm exp}(-\sum_{i<j}^{N}\alpha_{ij}({\bf r}_i-{\bf r}_j)^2)$.
These Gaussian functions are widely used in variational
calculations (see,
for example, \cite{temper,etbm}). We will apply a more general form of the
correlated
Gaussian functions which allow for nonzero orbital angular momentum, and
will use the more convenient Jacobi relative coordinates instead of
the relative distance vectors.
The correlated Gaussians have an important advantage. Their
Hamiltonian matrix elements can be analytically calculated
in a unified framework, thus enabling one to avoid the formidable
calculation involving the correlation factor $F$.
\par\indent
The variational approximation, however, may run into difficulties for the
following reasons:
(i) if the nonlinear parameters specifying the basis functions are varied,
it is difficult to optimize them,
 (ii) if they are not, then the number of terms required may be excessively
large, and, in both cases,
(iii) the trial function of proper symmetry becomes extremely involved. For
example, conventional methods \cite{crcbgv,temper}
for the choice of the Gaussian parameters lead to prohibitively large bases for
more than 3 or 4 particles, which has limited the applicability of the
Gaussian basis to few-body problems.
\par\indent
One can circumvent the optimization problem including large number of
nonlinear parameters or the diagonalization of huge matrices by using the
SVM. The SVM attempts to set up the most appropriate basis functions by the
following stepwise procedure: One generates a would-be
basis function by choosing the nonlinear parameters randomly, judges its
utility by the energy gained by including it in the basis, and
either keeps or discards it. One
repeats this ``trial and error'' procedure until the basis set up leads to
convergence. The original procedure of the SVM, proposed in \cite{SVM},
has recently been developed further and successfully applied to multicluster
descriptions of light exotic nuclei, such as $^6$He=$\alpha+ n + n$,
$^8$He=$\alpha+ n + n + n + n$, $^9$Li=$\alpha + t  + n + n$,
and $^9$C=$\alpha+^3$He$+p + p$ \cite{VSL,SVAO}. Learning from these
applications, we have now generalized and refined the method further to
encompass diverse systems emerging in nuclear and  atomic physics.
\par\indent
Besides the large number of nonlinear parameters, the treatment
of the increasing number of partial waves in the expansion of the wave
function would also pose a formidable task. We propose here an alternative
formulation to cope with this problem.
Instead of using the partial wave expansion, the angular dependence of the
wave function is represented by a single solid spherical harmonics
whose argument contains additional variational parameters. This form
makes the calculation of the matrix elements for nonzero orbital angular
momentum much simpler than other methods.
\par\indent
   It will be demonstrated that the present method has several unique
features: It is based on a fully analytical calculation for most types of
interactions and thus ensures high accuracy and speed. Its calculational
scheme is quite universal and needs no change depending on whether the
system contains nuclear or Coulombic or other
interactions. It has no  difficulty in
treating the system of particles of unequal masses. More importantly, the
wave function is obtained in a compact, analytical form and thereby can be
readily used in calculations of physical properties.
\par\indent
As you will see later, the present method has turned out to be
very accurate, and we think it is worth while to make the method and the
results
easily available and  reproducible for interested readers. We collect all
the needed ingredients of our method in order.
Some of the formulae are our original developments or generalizations
of known relations to $N$-particle matrix elements, and some others are
collected here to make the paper self-contained. The calculation of the
matrix elements presented here is  different from the one of Refs.
\cite{KA91,epep} in many aspects: The motion of the centre-of-mass is
removed from both the Hamiltonian and  the wave function. Two-particle
potential matrix elements
of arbitrary radial form factor are evaluated in a unified way by reducing
them to the calculation of appropriate correlation functions corresponding
to the interaction.
The calculation of the matrix elements is extended to nonzero
orbital angular momentum as well. The symmetrization postulate is imposed
on the wave function at the single-particle level, which provides several
advantages, especially in evaluating the matrix elements of state-dependent
realistic nuclear interactions.
\par\indent
The organization of the paper is as follows. Section II defines the
correlated Gaussian basis functions and gives the details of the stochastic
procedure of selecting the basis set. Section III contains the method of
calculating the matrix elements.  The main steps are the calculation of
matrix elements in Slater determinants (or permanents for bosons)
consisting of single-particle Gaussian wave packets, the elimination of the
center-of-mass motion with a very simple manipulation, and the
transformation to the correlated Gaussian basis. This section also presents
the modifications needed for treating systems of particles of unequal
masses. Section IV presents  numerical results for various systems of
particles which interact via nuclear potentials or power-law potentials.
Section V gives a brief summary. In the appendices the most important
auxiliary formulae are collected to facilitate any future use of the
formulation.

\section{The correlated Gaussians and the stochastic variational method}
\subsection{Basis functions}
Since the variational method is always limited by the form chosen as a
trial function, the trial
function must be flexible enough to be able to describe
the full variety of correlations between the nucleons, e.g., the short-range
correlation due to the strong repulsive force, the $\alpha$-clustering
typical in some light nuclei, or  the long-range
correlation at large distances in light halo nuclei. The correlation
between the nucleons can
be described by functions of appropriate relative coordinates.
\par\indent
   Any square-integrable function with angular momentum
$lm$ can be approximated, to any desired accuracy, by a linear combination
of nodeless harmonic-oscillator functions (Gaussians) of continuous size
parameter $a$:
\begin{equation}
\Gamma_{lm}({\bf r}) \sim {\rm e}^{-{\frac{1}{2}} a r^2}
{\cal Y}_{lm}({\bf r}),\ \ \ {\rm with}\ \ \ {\cal Y}_{lm}({\bf r})=r^l
Y_{lm}(\hat{\bf r}).
\end{equation}
A generalization
of this to $N$-nucleon systems contains a product of the Gaussians as
mentioned in the previous section. It is convenient to use a set of the
Jacobi coordinates ${\bf x}=({\bf x}_1,...,{\bf x}_{N-1})$, instead of
$N(N\!\!-\!\!1)/2$ relative distance vectors $({\bf r}_i-{\bf r}_j)$.
An $N$-nucleon basis function, a so-called correlated Gaussian,
then looks like
\begin{equation}
\psi_{(LS)JMTM_T}({\bf x}, A)={\cal A} \bigl\{
{\rm e}^{-{1 \over 2}\tilde{{\bf x}}A {\bf x}} \left[ \theta_{L}({\bf x})
 \chi_{S}\right]_{JM} {\cal X}_{TM_T}\bigr\},
\label{corrGauss}
\end{equation}
where $\tilde{{\bf x}}$, the transpose of ${\bf x}$, stands for the row
vector comprising the Jacobi coordinates. $\chi$ and ${\cal X}$ are the
spin and isospin functions.
$A$ is an $(N\!\!-\!\!1)\!\!\times\!\!(N\!\!-\!\!1)$ positive-definite,
symmetric matrix of nonlinear parameters, specific to each basis element,
and the quadratic form, $\tilde{\bf x}A{\bf x}$, involves scalar products
of the Cartesian vectors:
\begin{equation}
\tilde{{\bf x}} A {\bf x}=\sum_{i=1}^{N-1} \sum_{j=1}^{N-1}
A_{ij}
{\bf x}_i\cdot{\bf x}_j .
\end{equation}
The operator ${\cal A}$ is an antisymmetrizer defined by
\begin{equation}
{\cal A}={1 \over \sqrt{N!} }\sum_P^{N!} {\rm sign}(P) P,
\end{equation}
where the sum runs over all permutations of the $N$ nucleon indices and
sign($P$) stands for the parity of the permutation $P$. For a system of
identical bosons, the antisymmetrizer is to be replaced with
a symmetrizer. For a
general case ${\cal A}$ is to represent the operator that imposes the proper
symmetry on the wave function.
\par\indent
The function $\theta_{LM_L}({\bf x})$ in Eq. (2), which represents the
angular part of the wave function, is a generalization of
$\cal Y$ and can be chosen as a vector-coupled
product of solid spherical harmonics of the Jacobi coordinates
\begin{equation}
\theta_{LM_L}({\bf x})=\left[[[{\cal Y}_{l_1}({\bf x}_1){\cal Y}_{l_2}
({\bf x}_2)]_{L_{12}}
{\cal Y}_{l_3}({\bf x}_3)]_{L_{123}}
,...\right]_{LM_L} .
\end{equation}
Each relative motion has a definite angular
momentum in Eq. (5). It may be important, however, to include several sets of
angular
momenta $(l_1,l_2,...,l_{N-1};L_{12},L_{123},...)$ for a realistic
description.  The various possible
partial wave contributions increase the basis dimension; moreover, the
calculation of matrix elements for this choice of $\theta_{LM_L}({\bf x})$
becomes too complicated. This choice is apparently
inconvenient especially as the number of  nucleons increases.
To avoid this, we propose a different choice as the generalization of
$\cal Y$:
\begin{equation}
\theta_{LM_L}({\bf x})=\eta_{KLM_L}({\bf u},{\bf x})=v^{2K+L} Y_{LM_L}
({\hat{\bf v}}),\ \ \ \ {\rm with}\ \ \
{\bf v}=\sum_{i=1}^{N-1} u_i {\bf x}_i.
\end{equation}
Only the total orbital angular momentum appears in this expression and it
contains a parameter $\tilde{\bf u}=(u_1,...,u_{N-1})$. The vector {\bf u}
may be considered as a variational parameter and one may try to minimize
the energy functional with respect to it. It defines a linear combination
of the Jacobi coordinates, ${\bf v}$, and the wave function of the system is
expanded in terms of its angle $\hat{\bf v}$. The minimization amounts to
finding the most suitable angle or a linear combination of angles. The
factor of $v^{2K+L}$ plays an important role in improving the short-range
behavior of the wave function. A remarkable advantage of this form of
$\theta_{LM_L}({\bf x})$ is that the calculation of matrix elements becomes
much simpler than in the former case because the coupling of $(N\!-\!1)$
angular momenta is completely avoided.
\par\indent
The two forms of $\theta_{LM_L}({\bf x})$ are in fact closely related to
each other. Any of the functions of Eq. (5) may be expressed in terms of
a linear combination of the terms, $v^{2K+L} Y_{LM_L}
({\hat{\bf v}})$, by using some appropriate sets of $\bf u$
values provided that each term satisfies the condition
$2K\!+\!L{\le}l_1\!\!+\cdot\cdot\cdot+\!l_{N-1}$ and contains a monomial of
degree $l_1+\cdot\cdot\cdot+l_{N-1}\!-\!2K\!-\!L$ in the variables,
${{\bf x}_1}^2,...,{{\bf x}_{N-1}}^2$.  Therefore, if one can calculate
the matrix elements using $\theta_{LM_L}({\bf x})$ defined in Eq. (6), then
those with the previous form of $\theta_{LM_L}({\bf x})$ can be obtained
readily.
\par\indent
The correlated Gaussian basis with the function
$\theta_{LM_L}({\bf x})$ of Eq. (6) has parity $(-1)^L$. To construct
a function with parity $(-1)^{L+1}$, Eq. (6) must be slightly
generalized, e.g., to
\begin{equation}
\theta_{LM_L}({\bf x})=[\eta_{KL}({\bf u},{\bf x})\eta_{01}
({\bf u}',{\bf x})]_{LM_L}.
\end{equation}

\par\indent
To assure positive definiteness, the matrix $A$ in Eq. (2) is in general
expressed as $A =\tilde{G}A'G$, where $G$ is an $(N\!-\!1)\!\times\!
(N\!-\!1)$
orthogonal matrix containing $(N\!-\!1)(N\!-\!2)/2$
parameters and $A'$ is a diagonal matrix, $(A')_{ij}=a'_{i}
\delta_{ij}$, including $(N\!-\!1)$ positive parameters $a'_i$.  Although no
restriction on the parameters of the matrix $G$
is in principle necessary, it is advisable to avoid too many variables
if possible. The most naive choice would be to take $G$ as a
unit matrix, which is equivalent to using only a single set of the
Jacobi coordinates, and then to try to reach
convergence by including higher partial waves successively. Many examples
show \cite{crcbgv,SVAO}, however, that this does not work well because
the convergence is generally slow and moreover the computational cost of
using high partial waves is quite expensive.
\par\indent
The matrix $G$ can also be chosen as one of the rotation matrices that
connect the set of the Jacobi coordinates to other sets of independent
relative coordinates. Figures
1a-1f show all topologically different sets of independent relative
coordinates for a system of six identical particles. The set of
coordinates in Fig. 1a is what we call the set of the Jacobi coordinates
${\bf x}$. A correlation conforming to a specific set of  relative
coordinates $\tilde{\bf x}'=({\bf x}'_1,...,{\bf x}'_{N-1})$ can be most
efficiently described by tailoring the form of the basis function to this
set of relative coordinates, that is, by using the form,
exp$\lbrace -\frac{1}{2}\sum_{i=1}^{N-1}a'_i{\bf x}'_i\cdot{\bf x}'_i
\rbrace$. Since the coordinates ${\bf x}'$ can be obtained by an
appropriate rotation ${\cal R}$ of the Jacobi coordinates as ${\bf x}'=
{\cal R}{\bf x}$, the basis function of such type can be clearly
encompassed in the trial function of Eq. (2) by choosing $G={\cal R}$.
The correlated Gaussian basis thereby allows for
various correlations between the nucleons and different asymptotics at large
distances flexibly. Depending on the character of the problem a more general
choice of $G$ might be necessary.

\par\indent
By selecting a set of basis functions $\lbrace \psi_i; i=1,...,{\cal K}\rbrace$
[$\psi_i\equiv \psi_{(L_iS_i)JMTM_T}({\bf x}, A_i)$]
that adequately spans the state space, the wave function of the
$N$-nucleon system can be expanded as
\begin{equation}
\Psi=\sum_{i=1}^{\cal K} c_i \psi_i ,
\end{equation}
where $\tilde{\bf c}=(c_1,...,c_{\cal K})$ is the set of linear
variational parameters. The Ritz variational method defined by this trial
function
reduces to the generalized algebraic eigenvalue problem
\begin{equation}
{\cal H} {\bf c} = E {\cal N} {\bf c},
\end{equation}
where ${\cal H}$ and ${\cal N}$ are, respectively, the matrices of the
Hamiltonian and of the overlap
\begin{equation}
{\cal H}_{ij}=\langle \psi_i \vert H \vert \psi_j \rangle, \ \ {\rm and} \ \ \
\
{\cal N}_{ij}=\langle \psi_i \vert \psi_j \rangle \ \ \ \  (i,j=1,...,{\cal
K}).
\end{equation}

\subsection{Stochastic selection of parameters and solution of the
eigenvalue problem}
Since a linear combination of the correlated Gaussians forms a
dense set, there are different sets of $A$ that
represent the wave function equally well. This enables one to
select the most appropriate parameters randomly. We set up the basis
stepwise by choosing $A$ from a preset domain of the parameter space and
increase the basis dimension one by one.
In the first step we select a number of parameter sets $A$
randomly, and keep the one that gives the lowest energy.  Next we generate
a new random set and calculate the energy with this two-element basis.
As one more basis state always lowers the energy, we quantify its
``utility'' by the energy gained by including it in the
basis.  If
the energy gain is larger than a preset value, $\epsilon$, then we admit this
state to the basis, otherwise we discard it and try a new random candidate.
This is repeated until the energy converges.
The rate of convergence can be controlled by dynamically decreasing the
value of $\epsilon$ during the search.
This procedure is more advantageous than the earlier versions
\cite{SVM,VSL} and, although
not a full  optimization, results in very good and relatively small bases.
A similar procedure, called ``stochastic diagonalization'' has been used to
determine the smallest eigenvalue of extremely large matrices \cite{SD}.
\par\indent
To have an economical algorithm for setting up the basis by a trial and
error method,
one has to find an efficient way to solve the eigenvalue problem, Eq. (9).
The full diagonalization is rather time
consuming and in fact unnecessary because (i) in the $({\cal K}\!\!+\!\!1)$th
step of the procedure
we can use the result of the ${\cal K}$th step and (ii)  to judge the
usefulness
of a would-be basis
state, only the lowest eigenvalue is needed. Let us
assume that in the ${\cal K}$th step the
Hamiltonian matrix ${\cal H}$ is diagonalized; its eigenenergies are $E_1 \le
E_2 \le ...
 \le E_{\cal K}$ and its
corresponding normalized eigenfunctions are
${\Psi_1},...,{\Psi_{\cal K}}$. The eigenvalue problem
in the $({\cal K}\!\!+\!\!1)$th step takes the form
\begin{eqnarray}
& &\left(
\begin{array}{ccccc}
E_1    & 0 & ... & 0&   \langle {\Psi_1} \vert H
\vert \psi_{{\cal K}+1} \rangle \\
0        &    &    &    &                \\
\vdots &    &     &  \vdots  &      \vdots \\
0       &    & ... & E_{\cal K} &  \langle {\Psi_{\cal K}} \vert H \vert
\psi_{{\cal K}+1} \rangle \\
\langle {\psi_{{\cal K}+1}} \vert H \vert \Psi_{1} \rangle  & & ... &
\langle {\psi_{{\cal K}+1} }\vert H \vert  { \Psi_{\cal K}}\rangle
&
\langle {\psi_{{\cal K}+1} }\vert H \vert  { \psi_{{\cal K}+1}}\rangle \\
\end{array}
\right)
\left(
\begin{array}{c}
c_1 \\
\vdots \\
\vdots \\
c_{\cal K}  \\
c_{{\cal K}+1} \\
\end{array}
\right)
\nonumber \\
& &=E
\left(
\begin{array}{ccccc}
1    &0 & ... & 0 &    \langle {\Psi_1}  \vert \psi_{{\cal K}+1} \rangle \\
0       &    &    &       &                \\
\vdots &    &     &   \vdots    &      \vdots \\
0        &    & ... &   1  &  \langle {\Psi_{\cal K}}  \vert \psi_{{\cal K}+1}
\rangle \\
\langle \psi_{{\cal K}+1}  \vert {\Psi_{1}} \rangle  & & ... &
\langle { \psi_{{\cal K}+1} } \vert \Psi_{\cal K}\rangle
&
\langle { \psi_{{\cal K}+1} } \vert \psi_{{\cal K}+1}\rangle \\
\end{array}
\right)
\left(
\begin{array}{c}
c_1 \\
\vdots \\
\vdots \\
c_{\cal K}  \\
c_{{\cal K}+1}
\end{array}
\right).
\end{eqnarray}
By using the Gram-Schmidt orthogonalization method, that is, by defining
\begin{equation}
\vert {\bar\psi_{{\cal K}+1}} \rangle={\vert \psi_{{\cal K}+1}\rangle
-\sum_{i=1}^{\cal K} \vert {\Psi_{i} } \rangle \langle {\Psi_{i}} \vert
\psi_{{\cal K}+1} \rangle \over
( \langle \psi_{{\cal K}+1} \vert\psi_{{\cal K}+1} \rangle - \sum_{i=1}^{\cal
K}
\langle \psi_{{\cal K}+1} \vert{\Psi_{i}} \rangle
\langle{\Psi_{i}} \vert\psi_{{\cal K}+1} \rangle )^{1/2}} ,
\end {equation}
this generalized eigenvalue equation can be reduced to the conventional
form
\begin{equation}
\left(
\begin{array}{ccccc}
E_1    & 0 & ... & 0&   q_1 \\
0        &    &    &    &                \\
\vdots &    &     &    &      \vdots \\
0       &    & ... & E_{\cal K} &  q_{\cal K} \\
q_1  & & ... & q_{\cal K}  & a \\
\end{array}
\right)
\left(
\begin{array}{c}
c_1 \\
\vdots \\
\vdots \\
c_{\cal K}  \\
c_{{\cal K}+1} \\
\end{array}
\right)
=E
\left(
\begin{array}{c}
c_1 \\
\vdots \\
\vdots \\
c_{\cal K}  \\
c_{{\cal K}+1} \\
\end{array}
\right) ,
\end{equation}
where
\begin{equation}
q_i=\langle {\Psi_i} \vert H \vert {\bar\psi_{{\cal K}+1}} \rangle,
\ \ \ \ \ \
a=\langle  {\bar\psi_{{\cal K}+1}} \vert H \vert  {\bar\psi_{{\cal K}+1}}
\rangle .
\end{equation}
The eigenvalues are easily obtained
by finding the roots of the secular equation
\begin{equation}
\lambda(E)\equiv\prod_{i=1}^{\cal K} (E_{\cal K}-E)\left((a-E)-\sum_{j=1}^{\cal
K} {q_j^2\over E_j - E }
\right)=0 .
\end{equation}
This secular equation has $({\cal K}\!\!+\!\!1)$ roots $\lbrace
{E'}_i;i=1,...,{\cal K}\!\!+\!\!1\rbrace$
fulfilling the inequalities
${E'}_1\le E_1\le{E'}_2\le E_2\le ... \le E_{\cal K}\le {E'}_{{\cal K}+1}$.
The eigenvectors are readily obtained after substituting the eigenvalues
$\lbrace {E'}_i;i=1,...,{\cal K}\!\!+\!\!1\rbrace$ into Eq. (13).
Note that one has to determine only the lowest eigenvalue ${E'}_1$ for the
admittance criterion.

\section{Calculation of the matrix elements}
In this section we will give the details of the method of calculating the
matrix elements between
the basis function of Eq. (2). The calculation consists of three steps:
(A)The calculation of the matrix elements between the Slater
determinants of the Gaussian wave-packet single-particle functions, (B)A
transformation from the single-particle coordinate representation to the
relative and center-of-mass coordinate representation, (C)An integral
transformation  from the Gaussian
wave-packet functions to the correlated Gaussian basis. A procedure similar
to steps (A) and (B) was used to manipulate algebraically the
antisymmetrization operation and the transformation of the coordinates for
complex cluster systems \cite{suzuki}. In step (A) the
Slater determinant for the $N$-nucleon wave function is constructed by
distributing the nucleons at positions $({\bf s}_1,...,{\bf s}_N)$.
These position vectors serve as the generator coordinates. The Slater
determinant of the
Gaussian wave packets is often used in nuclear theory, e.g., in cluster
model \cite{suzuki,Brink,wild,supple} and fermionic or antisymmetrized
molecular dynamics
\cite{feldmeier,horiuchi}. The Hamiltonian matrix elements
are analytically evaluated with the use of technique of the Slater
determinants \cite{Lowden,Brink},
and can be expressed as a function of the generator coordinates. In step (B)
the center-of-mass motion is completely separated from the intrinsic motion,
and thus the trial wave function acquires the translational invariance. The
separation of the center-of-mass motion is particularly simple in this
formulation. In the last step (C) the matrix elements expressed in terms of
the intrinsic generator coordinates are transformed to those between the
correlated Gaussian basis functions with a definite angular momentum. Some
of the essential parts of the calculational scheme is our original
development, and some of them is a
generalization of the technique used in the nuclear cluster model
(see, for example, \cite{Kami}). We also show in subsection III.D those
modifications which are needed to treat the system of particles of
unequal masses.
\subsection{Slater determinants of Gaussian wave packets}
The $i$th nucleon with mass $m$, spin $\sigma_i$ and isospin
$\tau_i$ is to be put in the single-particle Gaussian wave packet
\begin{equation}
{\hat \varphi}_{{{\bf s}_i}{\sigma_i}{\tau_i}}^{\nu}({\bf r}_i)=
\varphi_{{\bf s}_i}^{\nu}({\bf r}_i)
\chi_{{1\over 2}\sigma_i}{\cal X}_{{1\over 2}\tau_i},
\end{equation}
with
\begin{equation}
\varphi_{{\bf s}_i}^{\nu}({\bf r}_i)=\left({2 \nu \over \pi}
\right)^{3/4} {\rm e}^{-\nu ({\bf r}_i-{\bf s}_i)^2},\ \ \ \ \ {\rm and}
\ \ \ \nu={m \omega \over 2 \hbar},
\end{equation}
where ${\bf r}_i$ is the position vector of the nucleon,
$\chi_{{1\over 2}\sigma_i}$ and ${\cal X}_{{1\over 2}\tau_i}$ are its spin
and isospin function.  The angular frequency $\omega$ is
not a variational parameter and may be taken an arbitrary constant. The
${\bf s}_i$ parameter or ``generator''
coordinate will be used in an integral transformation to derive the matrix
elements between the Gaussian basis functions.
A Slater determinant of these Gaussian packets is defined by
\begin{equation}
\phi_{\kappa}({\bf s}_1,...,{\bf s}_N)={\cal A} \Biggl\{
\prod_{i=1}^{N}{\hat \varphi}_{{{\bf s}_i}{\sigma_i}{\tau_i}}^{\nu}
({\bf r}_i) \Biggr\} ,
\end{equation}
where $\kappa=(\sigma_1\tau_1,...,\sigma_N\tau_N)$ is the set of the
spin-isospin quantum numbers of the
nucleons.
The spins and the isospins of the nucleons are successively coupled to
add up, respectively, to the
total spin $S M_S$ and isospin $T M_T$  of the $N$-nucleon system:
\begin{equation}
{\chi}_{SM_S}=
\left[[[\chi_{1\over 2}\chi_{1\over 2}]_{S_{12}}
\chi_{1\over 2} ]_{S_{123}}
,...\right]_{SM_S} ,\ \ \ \
{\cal X}_{TM_T}=
\left[[[{\cal X}_{1\over 2}{\cal X}_{1\over 2}]_{T_{12}}
{\cal X}_{1\over 2}]_{T_{123}}
,...\right]_{TM_T} .
\end{equation}
To simplify the notation, the intermediate quantum numbers are suppressed
in the following.
The wave function in the ``generator coordinate space'' with the
definite spin and isospin quantum numbers is a linear
combination of the Slater determinants of the Gaussian packets:
\begin{equation}
\Phi_{SM_STT_z}({\bf s}_1,...,{\bf s}_N)=
{\cal A} \lbrace
\varphi_{{\bf s}_1}^{\nu}({\bf r}_1)...
\varphi_{{\bf s}_N}^{\nu}({\bf r}_N) {\chi}_{SM_S}{\cal X}_{TM_T}\rbrace
=\sum_{\kappa} c_{\kappa} \phi_{\kappa}({\bf s}_1,...,{\bf s}_N) ,
\end{equation}
where $c_{\kappa}$ is a product of the Clebsch-Gordan coefficients needed
to couple the spin and isospin as defined in Eq. (19).
\par\indent
The Hamiltonian of the $N$-nucleon system reads as
\begin{equation}
H=\sum_{i=1}^{N} {{\bf p}_i^2 \over 2 m} +\sum_{i<j}^{N} V_{ij} .
\end{equation}
The matrix elements of the Slater determinants can easily be evaluated using
the well-known rules \cite{Lowden,Brink}. To make this paper
self-contained, we have collected all the needed ingredients in Appendices
A, B and C. The overlap of the Slater determinants is found to take the form
\begin{equation}
\langle
\Phi_{SM_STM_T}({\bf s}_1,...,{\bf s}_N)
\vert
\Phi_{SM_STM_T}({{\bf s}'}_1,...,{{\bf s}'}_N)
\rangle=
\sum_{i=1}^{n_{o}} C_{i}^{(o)}
{\rm e}^{-{1\over 2} \tilde{{\bf s}} A_{i}^{(o)} {\bf s}}  ,
\end{equation}
where $A_{i}^{(o)}$ is a $2N\!\times\!2N$ real, symmetric matrix and
$\tilde{{\bf s}}$ stands for the $2N$-dimensional row vector
comprising the single-particle generator coordinates,
$({\bf s}_1,...,{\bf s}_N,{\bf s}'_1,...,{\bf s}'_N)$.
To simplify the notation, we refer to the set of the vectors
$({\bf s}'_1,...,{\bf s}'_N)$ alternatively as
$({\bf s}_{N+1},...,{\bf s}_{2N})$. Note, therefore, that
the quadratic form, $\tilde{{\bf s}} A_{i}^{(o)} {\bf s}$, reads as
\begin{equation}
\tilde{{\bf s}} A_{i}^{(o)} {\bf s}=\sum_{j=1}^{2N} \sum_{k=1}^{2N}
\left(A_{i}^{(o)}\right)_{jk}
{\bf s}_j\cdot{\bf s}_k .
\end{equation}
The matrix elements of the kinetic energy
operator can also be expressed in terms of the same $C_{i}^{(o)}$'s and
$A_{i}^{(o)}$'s as
\begin{eqnarray}
& & \langle
\Phi_{SM_STM_T}({\bf s}_1,...,{\bf s}_N)
\vert \sum_{i=1}^{N} {{\bf p}_i^2 \over 2 m}  \vert
\Phi_{SM_STM_T}({{\bf s}'}_1,...,{{\bf s}'}_N)
\rangle
\nonumber \\
& &={\hbar \omega \over 2}
\sum_{i}^{n_o} C_{i}^{(o)} \left({3 \over 2}N-{1\over 2}
\tilde{{\bf s}} A_{i}^{(o)} {\bf s} \right)
{\rm e}^{-{1\over 2} \tilde{{\bf s}} A_{i}^{(o)} {\bf s}}.
\end{eqnarray}
The matrix elements of any term of the two-body interaction
can be expressed as an integral of the two-particle correlation function
multiplied by the radial form factor, $V(r)$, of the term $V_{ij}$ as below:
\begin{eqnarray}
& &\langle
\Phi_{SM_STM_T}({\bf s}_1,...,{\bf s}_N)
\vert \sum_{i<j}^{N} V_{ij}  \vert
\Phi_{S'M_S'T'M_T'}({{\bf s}'}_1,...,{{\bf s}'}_N)
\rangle
\nonumber \\
& & =\int d{\bf r} V(r) {\rm e}^{-\nu r^2}
\sum_{i=1}^{n_{p}} C_{i}^{(p)}
P_{i}({\bf s},{\bf r})
{\rm e}^{-{1\over 2} {\tilde{\bf s}} A_{i}^{(p)} {\bf s}+{\bf d}_i \cdot
 {\bf r}},
\end{eqnarray}
where
$P_{i}({\bf s},{\bf r})$ is a polynomial of ${\bf s}$ and ${\bf r}$,
$A_{i}^{(p)}$ a $2N\!\times\!2N$ symmetric matrix, and
${\bf d}_i\cdot{\bf r}$ takes the form
\begin{equation}
{\bf d}_i\cdot{\bf r}=\sum_{j=1}^{2N} {\cal D}_{(i)j}{\bf s}_j
\cdot{\bf r}.
\end{equation}
The polynomial part reduces to unity
($P_{i}({\bf s},{\bf r})=1$) in the case of pure central forces,
but it has a rather simple form for spin-orbit, tensor, and other
interactions as well. See Appendices B and C. The $C_{i}^{(o)}$'s,
$A_{i}^{(o)}$'s,
$C_{i}^{(p)}$'s, and $A_{i}^{(p)}$'s, etc. are obtained with the use of
mathematical manipulation languages or fortran programs. See Appendix C.

\subsection{Transformation to relative and center-of-mass coordinates}
To eliminate the center-of-mass motion, we transform the single-particle
coordinates to the relative and center-of-mass coordinates. For this purpose
we choose one particular set of relative coordinates, the Jacobi
coordinates, which is expressed in terms of the single-particle
coordinates ${\bf r}_i$ as
\begin{equation}
{\bf x}_i=\sum_{k=1}^{N} U_{ik} {\bf r}_k  \ \ \ \ (i=1,...,N),
\end{equation}
where the transformation matrix $U$ is defined by
\begin{equation}
U=
\left(
\begin{array}{ccccc}
-1            &     1     &    0 &... &  0              \\
-{1 \over 2}  & -{1 \over 2} &    1 &... &  0          \\
\vdots          &                &      &    &\vdots         \\
-{1 \over N-1}&-{1\over N-1} & ...  &... & 1             \\
{1\over N}    &{1\over N}    & ...  &... &{1 \over N}
\end{array}
\right),
\ \ \ \ \ \ {\rm and}\ \ \ \ \
U^{-1}=
\left(
\begin{array}{ccccc}
-{1 \over 2} & -{1 \over 3}  &... & -{1 \over N} & 1     \\
 {1 \over 2}  & -{1 \over 3} &     & -{1 \over N} & 1     \\
 0     & {2 \over 3} &      & \vdots    & \vdots       \\
\vdots     & \vdots      &      &  \vdots   & \vdots    \\
  0  &  0  & ...  & {N-1 \over N} &  1
\end{array}
\right).
\end{equation}
Similarly, the single-particle generator coordinates are transformed to
the relative and center-of-mass generator coordinates:
\begin{equation}
{\bf S}_i=\sum_{k=1}^{N} U_{ik} {\bf s}_k,  \ \ \ \ \ \
{\bf S}_i'=\sum_{k=1}^{N} U_{ik} {\bf s}_k'  \ \ \ \ \ \ (i=1,...,N).
\end{equation}
The reduced masses corresponding to the transformation $U$ are given by
\begin{equation}
\mu_i={i \over i+1} m  \ \ \ \  (i=1,...,N\!\!-\!\!1),
\ \ \ \ \ \ \ {\rm and}
\ \ \ \ \ \mu_N=N m.
\end{equation}
The product of the Gaussian single-particle
wave packets can then be written as a product of Gaussians depending on the
relative and center-of-mass coordinates:
\begin{equation}
\prod_{i=1}^{N} \varphi_{{\bf s}_i}^{\nu}({\bf r}_i)=
\prod_{i=1}^{N} \varphi_{{\bf S}_i}^{\gamma_i}({\bf x}_i) ,
\end{equation}
with
\begin{equation}
\gamma_i={\mu_i \omega \over 2 \hbar}.
\end{equation}
By using Eq. (31) and noting that the last factor of the product depends
only on the
center-of-mass coordinate, which is symmetric under the exchange of nucleons,
the $N$-nucleon wave function can be rewritten as
\begin{equation}
\Phi_{SM_STM_T}({\bf s}_1,...,{\bf s}_N)=
\Psi_{SM_STM_T}({\bf S}_1,...,{\bf S}_{N-1})
\varphi_{{\bf S}_N}^{\gamma_N}({\bf x}_N) ,
\end{equation}
which defines the intrinsic function that depends solely
on the relative coordinates,
\begin{equation}
\Psi_{SM_STM_T}({\bf S}_1,...,{\bf S}_{N-1})=
{\cal A} \lbrace
\varphi_{{\bf S}_1}^{\gamma_1}({\bf x}_1) ,...,
\varphi_{{\bf S}_{N-1}}^{\gamma_{N-1}}({\bf x}_{N-1})
{\chi}_{SM_S}{\cal X}_{TM_T}\rbrace .
\end{equation}
As will be shown in the next subsection, this function serves as a
generating function of the correlated Gaussian basis.
\par\indent
The Hamiltonian of Eq. (21) can be recast to the relative plus
center-of-mass terms as
\begin{equation}
H=\left( \sum_{i=1}^{N-1} {{{\bf P}_i}^2 \over 2 \mu_i}+
\sum_{i<j}^{N} V_{ij}
\right)+{{{\bf P}_N}^2 \over 2 \mu_N}\ \  {\equiv}\ \  H_{rel}+T_{cm} ,
\end{equation}
where ${\bf P}_i$ is the momentum canonically conjugate to the Jacobi
coordinate ${\bf x}_i$. The matrix elements of the Hamiltonian are then
\begin{eqnarray}
& & \langle \Phi_{SM_STM_T}({\bf s}_1,...,{\bf s}_N)
\vert  H  \vert \Phi_{S'M_S'T'M_T'} ({{\bf s}'}_1,...,{{\bf s}'}_N) \rangle
 \nonumber \\
& &=\langle \Psi_{SM_STM_T}({\bf S}_1,...,{\bf S}_{N-1})
\vert  H_{rel} \vert
\Psi_{S'M_S'T'M_T'}({{\bf S}'}_1,...,{{\bf S}'}_{N-1}) \rangle
\langle\varphi_{{\bf S}_N}^{\gamma_N} \vert
\varphi_{{{\bf S}'}_N}^{\gamma_N}\rangle   \nonumber\\
& &
+\langle \Psi_{SM_STM_T}({\bf S}_1,...,{\bf S}_{N-1}) \vert
\Psi_{S'M_S'T'M_T'}({{\bf S}'}_1,...,{{\bf S}'}_{N-1}) \rangle
\langle\varphi_{{\bf S}_N}^{\gamma_N} \vert
T_{cm} \vert
\varphi_{{{\bf S}'}_N}^{\gamma_N}\rangle  .
\end{eqnarray}
The matrix elements of the Hamiltonian $H_{rel}$ can be expressed by
integrating
over ${\bf S}_N$
\begin{eqnarray}
& &\langle \Psi_{SM_STM_T}({\bf S}_1,...,{\bf S}_{N-1}) \vert  H_{rel} \vert
\Psi_{S'M_S'T'M_T'}({{\bf S}'}_1,...,{{\bf S}'}_{N-1}) \rangle
 \nonumber \\
& & = \left({{\gamma_N} \over 2\pi}\right)^{3/2}
\int d{\bf S}_N
\langle \Phi_{SM_STM_T}({\bf s}_1,...,{\bf s}_N)
\vert  H  \vert
\Phi_{S'M_S'T'M_T'} ({{\bf s}'}_1,...,{{\bf s}'}_N) \rangle ,
\end{eqnarray}
where use has been made of the single-particle matrix elements in
Appendix A and the formula
\begin{equation}
\int d{\bf S}_N \left(-{{\gamma_N} \over 2}({\bf S}_N -{{\bf S}'}_N)^2
\right)^n
{\rm e}^{-{{\gamma_N} \over 2}({\bf S}_N -{{\bf S}'}_N)^2}=
\left(-{1\over 2}\right)^{n}(2n+1)!! \left({2\pi \over
{\gamma_N}}\right)^{3\over2}.
\end{equation}
\par\indent
Equation (37) shows that the matrix elements between the intrinsic
function can be obtained in a simple way by factorizing
the ${\bf S}_N$-dependent terms from the matrix elements in the
single-particle basis:
The quadratic forms in Eqs. (22), (24) and (25), with the help of
Eq. (36), should take the form
\begin{equation}
\tilde{{\bf s}} A_i^{(k)} {\bf s}=\tilde{{\bf S}} B_i^{(k)} {\bf S}+
{\gamma_N} ({\bf S}_N-{{\bf S}'}_N)^2  \ \ \ \ (k=o,p).
\end{equation}
Here the $(2N-2)$-element column vector  ${\bf S}$ is an abbreviation for the
set of the Jacobi generator-coordinate vectors
$({\bf S}_1,...,{\bf S}_{N-1},{{\bf S}'}_1,...,{{\bf S}'}_{N-1})$.
 The matrix $B_i^{(k)}$ is the $(2N\!\!-\!\!2)\!\!\times\!\!(2N\!\!-\!\!2)$
symmetric
matrix defined by dropping
the $N$th and $2N$th rows and columns of the matrix $\tilde{T} A_i^{(k)} T$,
where
\begin{equation}
T=
\left(
\begin{array}{cc}
U^{-1} & 0 \\
0 & U^{-1}
\end{array}
\right) .
\end{equation}
It is clear from Eq. (36) that the polynomials $P_{i}({\bf s},{\bf r})$
and the vectors ${\bf d}_i$
defined in Eq. (25) can depend only on the relative generator coordinates.
The dependence of the matrix elements on the center-of-mass variables,
${\bf S}_N$ and ${\bf S}'_N$, can, therefore, be factorized and the
integration in Eq. (37) reduces to such a simple form as the one in Eq. (38).
After the integration over ${\bf S}_N$ the matrix elements can be
expressed in terms of the Jacobi generator coordinates in the form:
\begin{equation}
\langle
\Psi_{SM_STM_T}({\bf S}_1,...,{\bf S}_{N-1})
\vert
\Psi_{SM_STM_T}({{\bf S}'}_1,...,{{\bf S}'}_{N-1})
\rangle=
\sum_{i=1}^{n_{o}} C_{i}^{(o)}
{\rm e}^{-{1\over 2} \tilde{\bf S} B_{i}^{(o)} {\bf S}},
\end{equation}
\begin{eqnarray}
& & \langle
\Psi_{SM_STM_T}({\bf S}_1,...,{\bf S}_{N-1})
\vert \sum_{i=1}^{N} {{\bf P}_i^2 \over 2 \mu_i}  \vert
\Psi_{SM_STM_T}({{\bf S}'}_1,...,{{\bf S}'}_{N-1})
\rangle
\nonumber \\
& &={\hbar \omega \over 2}
\sum_{i}^{n_o} C_{i}^{(o)} \left({3 \over 2}(N-1)-{1\over 2}
\tilde{\bf S} B_{i}^{(o)} {\bf S}\right)
{\rm e}^{-{1\over 2} \tilde{\bf S} B_{i}^{(o)} {\bf S}},
\end{eqnarray}
\begin{eqnarray}
& &\langle
\Psi_{SM_STM_T}({\bf S}_1,...,{\bf S}_{N-1})
\vert \sum_{i<j}^{N} V_{ij}  \vert
\Psi_{S'M_S'T'M_T'}({{\bf S}'}_1,...,{{\bf S}'}_{N-1})
\rangle
\nonumber \\
& & =\int d{\bf r} V(r) {\rm e}^{-\nu r^2}
\sum_{i=1}^{n_{p}} C_{i}^{(p)}
P_{i}({\bf S},{\bf r})
{\rm e}^{-{1\over 2} \tilde{\bf S} B_{i}^{(p)} {\bf S}+{\bf D}_i
\cdot {\bf r}}.
\end{eqnarray}
Here the convention of renumbering the set of the vectors
$({{\bf S}'}_1,...,{{\bf S}'}_{N-1})$ of the ket as $({\bf S}_N,...,
{\bf S}_{2N-2})$ is used to simplify the notation and thus ${\bf S}$ stands
for the set of the vectors $({\bf S}_1,...,{\bf S}_{2N-2})$. The vector
${\bf D}_i$ is given by $\sum_{j=1}^{2N-2} {\hat{D}}_{(i)j}
 {\bf S}_j $, where the ${\hat D}_{(i)j}$'s are formed from the elements
of the row vector
$\tilde{\cal D}_{(i)}T$ by omitting
its $N$th and $2N$th columns. The column vector
${\cal D}_{(i)}$ is defined in Eq. (26).

\subsection{Integral transformation to the correlated Gaussian basis}
\par\indent
Here we show how to evaluate the matrix elements in the correlated Gaussian
basis. Let us choose the correlated Gaussians with the form of Eq. (6) and
introduce the function
\begin{equation}
f_{KLM}({\bf u},{\bf x},A)=\eta_{KLM} ({\bf u},{\bf x}){\rm e}^{-{1\over 2}
\tilde{\bf x} A {\bf x}},
\end{equation}
where
\begin{equation}
\eta_{KLM}({\bf u},{\bf x})=v^{2K+L} Y_{LM}({\hat{\bf v}}),\ \ \ \ {\rm with}
 \ \ \ \ \ {\bf v}=\sum_{i=1}^{N-1} u_i {\bf x}_i=\tilde{\bf u}{\bf x},
\end{equation}
and where $\tilde{\bf u}=(u_1,...,u_{N-1})$. Note that $\bf u$ is a set
of $(N\!-\!1)$ real numbers, whereas $\bf x$ are the $(N\!-\!1)$
Jacobi coordinates. The calculation of the matrix elements
becomes simple if one uses a generating function of the correlated Gaussian.
In fact, the following function $g$ is found to
be most convenient to generate the function $f$:
\begin{equation}
f_{KLM}({\bf u},{\bf x},A)={1\over B_{KL}}
\int d{\hat {\bf t}} Y_{LM}({\hat {\bf t}})
\left( {d^{2K+L} \over d\alpha^{2K+L}}
g(\alpha,{\bf t};{\bf u},{\bf x},A)
\right)_{\alpha=0 \atop t=|{\bf t}|=1},
\end{equation}
where
\begin{equation}
g(\alpha,{\bf t};{\bf u},{\bf x},A)=
{\rm e}^{-{1\over 2}\tilde{\bf x} A {\bf x} +\alpha {\bf v}\cdot{\bf t}},
\end{equation}
\begin{equation}
B_{nl}={4 \pi (2n+l)! \over 2^n n! (2n+2l+1)!!}.
\end{equation}
Equation (46) is easily proved by using the simple formula
\begin{equation}
({\bf v}\cdot{\bf t})^{k}=
v^k t^k \sum_{n,l \ge 0 \atop 2n+l=k} B_{nl} \sum_{m=-l}^{l}
Y_{lm}({\hat {\bf v}})
Y_{lm}({\hat {\bf t}})^{\ast}.
\end{equation}
For a case where the function $\theta_{LM_L}({\bf x})$ of Eq. (7) is needed,
the generating function $g$ of Eq. (47) must be generalized to include
another factor $\alpha'{\bf v}'\cdot{\bf t}'$ . Since the following derivation
remains essentially unaffected by this generalization, we will assume
Eq. (6) as $\theta_{LM_L}({\bf x})$.
\par\indent
The generating function $g$ can be related to the product of the Gaussians
centered around $\{{\bf S}_i;i=1,...,N\!-\!1\}$ through an integral
transformation. To show this, we express the product of the Gaussian
wave packets as
\begin{equation}
\prod_{i=1}^{N-1}
\varphi_{{\bf S}_i}^{\gamma_i}({\bf x}_i)=\left(\frac{{\rm det}\Gamma}{\pi^
{N-1}}\right)^{3/4}{\rm e}^{-\frac{1}{2}\tilde{\bf x}\Gamma{\bf x}+
\tilde{\bf x}
\Gamma{{\bf S}_H}-\frac{1}{2}{{\tilde{\bf S}}_H}
\Gamma{{\bf S}_H}}
\end{equation}
with an $(N\!\!-\!\!1)\!\times\!(N\!\!-\!\!1)$ diagonal matrix
\begin{eqnarray}
\Gamma=
\left(
\begin{array}{cccc}
 2\gamma_1 & 0 & ...&  0       \\
 0  & 2\gamma_2 &    &  \vdots    \\
\vdots   &       &      &  \vdots         \\
 0  &...  & ...  & 2\gamma_{N-1}
\end{array}
\right),
\end{eqnarray}
where ${{\bf S}_H}$ stands for the set of the generator coordinate
vectors
$({\bf S}_1,...,{\bf S}_{N-1})$.
 By using the
familiar formula of the $n$-dimensional Gaussian integration
\begin{equation}
\int d{\bf x} {\rm e}^{-{1\over 2}\tilde{\bf x}A {\bf x}+\tilde{\bf T}
{\bf x}}=
\left({(2\pi)^{n} \over {\rm det} A }\right)^{3/2}
{\rm e}^{{1\over 2}\tilde{\bf T} A^{-1} {\bf T}},
\end{equation}
it is easy to prove the following equation by a direct calculation
\begin{equation}
g(\alpha,{\bf t};{\bf u},{\bf x},A)= \left(\frac{({\rm det}\Gamma)^{3/2}}
{(4\pi)^{(N-1)/2}{\rm det}(\Gamma-A)}
\right)^{3/2}
{\rm e}^{-{1\over 2} \tilde{\bf T} C {\bf T}}
\int d{{\bf S}_H}
{\rm e}^{-{1\over 2} {{\tilde{\bf S}}_H} Q
{{{\bf S}_H}}+\tilde{\bf T}{{\bf S}_H}}
\left(\prod_{i=1}^{N-1}
\varphi_{{\bf S}_i}^{\gamma_i}({\bf x}_i)
\right),
\end{equation}
where $\tilde{\bf T}=({\bf T}_1,...,{\bf T}_{N-1})$, and
\begin{equation}
C=\Gamma^{-1}(\Gamma-A)\Gamma^{-1} ,\ \ \ \ \
Q=C^{-1}-\Gamma ,
\end{equation}
\begin{equation}
{\bf T}_i=\alpha {\bf t}\sum_{j=1}^{N-1}(\Gamma C)^{-1}_{ij}
u_j \ \ \ \ (i=1,...,N\!\!-\!\!1).
\end{equation}
By combining Eqs. (46) and (53), the correlated Gaussian basis is found to
be generated from the intrinsic state given in Eq. (34) by the integral
transformation
\begin{eqnarray}
& & {\cal A} \{f_{KLM}({\bf u},{\bf x},A){\chi}_{SM_S}{\cal X}_{TM_T}\}
={1\over B_{KL} }\left(\frac{({\rm det}\Gamma)^{3/2}}
{(4\pi)^{(N-1)/2}{\rm det}(\Gamma-A)}
\right)^{3/2}
\nonumber \\
&\times& \int d{\hat {\bf t}}
Y_{LM}({\hat {\bf t}})
\left( {d^{2K+L} \over d\alpha^{2K+L}}
{\rm e}^{-{1\over 2} \tilde{\bf T} C
{\bf T}}
\int d{{{\bf S}_H}}
{\rm e}^{-{1\over 2} {{\tilde{\bf S}}_H} Q
{{{\bf S}_H}}+\tilde{\bf T}{{{\bf S}_H}}}
\Psi_{SM_STM_T}({\bf S}_1,...,{\bf S}_{N-1})
\right)_{\alpha=0 \atop t=1} .
\end{eqnarray}
\par\indent
The matrix elements between the correlated Gaussians are now easily
obtained
by the integral transformation from those expressed in terms of the relative
generator coordinates ${\bf S}$. Using Eq. (56) gives a general formula to
calculate a matrix element for any translation-invariant operator
$\cal O$
\begin{eqnarray}
& &\langle {\cal A} \{f_{KLM}({\bf u},{\bf x},A){\chi}_{SM_S}{\cal X}_{TM_T}\}
\vert {\cal O} \vert {\cal A'} \{f_{K'L'M'}({\bf u}',{\bf x},A')
{\chi}_{S'M'_S}{\cal X}_{T'M'_T}\}\rangle
\nonumber \\
&=&\frac{1}{B_{KL}B'_{K'L'}}\left(\frac{({\rm det}\Gamma)^3}
{(4\pi)^{(N-1)}{\rm det}(\Gamma-A){\rm det}(\Gamma-A')} \right)^{3/2}
\nonumber \\
&\times& \int\!\int d{\hat {\bf t}}d{\hat {\bf t}}'
Y_{LM}({\hat {\bf t}})^{*} Y_{L'M'}({\hat {\bf t}}')
\left( {d^{2K+L+2K'+L'} \over d\alpha^{2K+L}d\alpha'^{2K'+L'}}
{\rm e}^{-{1\over 2} \tilde{\bf T} C {\bf T}} \right.\\
&\times& \left. \int d{{\bf S}}
{\rm e}^{-{1\over 2} \tilde{{\bf S}} Q {{\bf S}}+\tilde{\bf T}{{\bf S}}}
\langle \Psi_{SM_STM_T}({\bf S}_1,...,{\bf S}_{N-1}) \vert {\cal O} \vert
\Psi_{S'M'_ST'M'_T}({{\bf S}'}_1,...,{{\bf S}'}_{N-1}) \rangle
\right)_{\alpha=\alpha'=0 \atop t=t'=1},
\nonumber
\end{eqnarray}
where $\tilde{\bf S}=({\bf S}_1,...,{\bf S}_{N-1},{{\bf S}'}_1={\bf S}_N,...,
{{\bf S}'}_{N-1}={\bf S}_{2N-2})$ and the matrices, $C$ and $Q$, and the
vectors ${\bf T}$ in Eq. (57), although the same notation is used as in
Eqs. (54) and (55), are extended to include the corresponding primed
quantities of the ket, that is,
\begin{equation}
C \longrightarrow \pmatrix{
     C & 0 \cr
     0 & C' \cr
    },
\ \ \ \
  Q \longrightarrow \pmatrix{
    Q & 0 \cr
    0 & Q' \cr
   },
\ \ \ \
 {\bf T} \longrightarrow \pmatrix{
    {\bf T} \cr
    {\bf T}' \cr
   }.
\end{equation}
As is shown in Eqs. (41)-(43), the ${\bf S}$-dependence of the matrix
elements is rather simple and the integration over ${\bf S}$ is done
analytically. Since the variables $\alpha, \alpha', {\bf t}$, and
${\bf t}'$ appear only through the vector ${\bf T}$, those operations
with respect to them as implied in Eq.
(57) are performed systematically. An illustrative example is given in
Appendix D. The coupling of the orbital and spin angular momenta
causes no difficulty. It is very satisfactory aesthetically that  matrix
elements between the basis functions with any sets of the relative
coordinates can be evaluated in a unified framework without
 any extra transformation of the coordinates. The choice of the set
of the relative coordinates amounts to the choice of the matrix $A$.

\subsection{Extension to the system of particles of unequal masses}
\par\indent
In this subsection we remark the modifications needed to treat
few-particle systems containing particles of unequal masses. As you will
see, most of the formulation presented in subsections III.A$-$III.C remains
unchanged.
Suppose that the masses of the $N$ particles are $m_1,m_2,...,m_N$. The
width parameter of the Gaussian wave packet is to be changed to
\begin{equation}
\nu_i={m_i\omega \over {2\hbar}} \ \ \ \ (i=1,...,N).
\end{equation}
The overlap and Hamiltonian matrix elements are obtained as a function of
the generator coordinates, $({\bf s}_1,...,{\bf s}_N,{\bf s}'_1,...,
{\bf s}'_N)$, in a form similar to the previous case. In this general case
of unequal masses,
the matrix $U$ in Eq. (28) which defines a set of the Jacobi coordinates
must be generalized to
\begin{equation}
U=
\left(
\begin{array}{ccccc}
-1            &     1     &    0 &... &  0              \\
-{m_1 \over m_{12}}  & -{m_2 \over m_{12}} &    1 &... &  0          \\
\vdots          &                &      &    &\vdots         \\
-{m_1 \over m_{12\cdot\cdot\cdot{N-1}}} &-{m_2 \over
m_{12\cdot\cdot\cdot{N-1}}}
& ...  &... & 1             \\
{m_1 \over m_{12\cdot\cdot\cdot{N}}} & {m_2\over m_{12\cdot\cdot\cdot{N}}}
 & ...  &... &{m_N \over m_{12\cdot\cdot\cdot{N}}}
\end{array}
\right),
\end{equation}
where $m_{12\cdot\cdot\cdot{i}}=m_1\!+\!m_2\!+\cdot\cdot\cdot+\!m_i$. The
reduced
masses corresponding to this $U$ are accordingly given, instead of
Eq. (30), by
\begin{equation}
\mu_i={m_{i+1}m_{12\cdot\cdot\cdot{i}} \over m_{12\cdot\cdot\cdot{i+1}}}
 \ \ \ \   (i=1,...,N\!\!-\!\!1),
\ \ \ \ \ \ \ {\rm and}
\ \ \ \ \ \mu_N=m_{12\cdot\cdot\cdot{N}}.
\end{equation}
What is important is to realize that Eq. (31), most crucial in eliminating
the center-of-mass motion, still holds even with these modifications as
\begin{equation}
\prod_{i=1}^{N} \varphi_{{\bf s}_i}^{\nu_i}({\bf r}_i)=
\prod_{i=1}^{N} \varphi_{{\bf S}_i}^{\gamma_i}({\bf x}_i).
\end{equation}
It is then easy to see that the rest of all the formulae are exactly the
same as the case of equal masses. We can conclude that the needed
modifications noted above are rather trivial and simple but still assures
the elimination of the center-of-mass motion. As a simple example of unequal
masses, the system of $t+d+\mu^{-}-$molecule will be considered in
subsection IV.B.

\section{Numerical results}
This section is devoted to present the solutions of various
$N\!\!=\!\!2-7$-body
problems by applying the method described above. To test the method,
different potentials (Yukawa, Gauss and Coulomb) have been used for bosonic
and fermionic systems. Some of the examples shown here has its own physical
significance, and
some other solutions may be considered as benchmark test and might
be useful in comparison of various few-body methods. One can expect that,
besides  the VMC\cite{mcv,vmc} and GFMC\cite{gfma,gfmb}, other methods
will also be extended to treat more than $N=4$-particle systems. As
only a few solutions are at present available for simple
potentials, the examples listed here may help to test other methods.
\par\indent
As was discussed in subsection II.A, there is no restriction on the choice
of the orthogonal matrix $G$. We have found, however, that those
special rotation matrices which connect different sets of the relative
coordinates especially suitable (see
also \cite{SVAO}) and will use them in what follows.
This greatly helps to reduce the number of  parameters of $G$.
In the stochastic selection of the basis elements,
these special matrices $G$ and the parameters of the diagonal matrix $A'$
are randomly chosen.
The vector ${\bf u}$ in Eq. (6) is also a variational
parameter. To avoid an excessively large number of variational parameters we
limited ${\bf u}$-vector values to those which are needed to
generate the function ${\theta}_{LM_{L}}({\bf x})$ of Eq. (5) for a given
set of angular momenta. Comparison of our calculation with others confirms
that these limitations have not deteriorated the accuracy of the present
calculation, that is, our trial function is flexible enough. In the
calculation the sets of angular momenta (i.e., the sets of these special
${\bf u}$ vectors) are also randomly chosen. The main advantage of using
${\bf u}$ lies in the simple and systematic evaluation of the matrix
elements from the point of view of both analytical and numerical
calculations. Further test calculations will be needed to explore the
utility of ${\bf u}$ as a variational parameter.
\par\indent
Because the dependence of the matrix elements on the variational parameters
is known as is shown in Appendix D, one can organize the numerical
calculation involved in the
random search economically. A change of the values of the parameters does
not require a recalculation of the whole matrix element. Once they have been
calculated for one set of values, to calculate them for many more requires
virtually no time. The average computational time
is 10 minutes for a four-body and 2 hours for a six-body calculation
on the VPP500 computer of RIKEN.

\subsection{Few-nucleon systems}
We have performed model calculations adopting different central potentials
as nucleon-nucleon interaction. Some of these model problems have already
been solved to high accuracy by various methods and therefore we can
directly compare the solutions. The potentials used for comparison of
different methods are
(i) the Malfliet-Tjon (MT-V) potential \cite{mt}, which has been most
extensively used as benchmark test in few-body
calculations, (ii) the Volkov ``super-soft'' core potential \cite{Volkov},
(iii) the Afnan-Tang S3 (ATS3) potential \cite{ATS3}
which exhibits a strong repulsive core and incorporates
a difference between the spin-singlet and spin-triplet channels, (iv) and
the Minnesota potential \cite{minesota} which reproduces
the most important low-energy nucleon-nucleon phase shifts. The Volkov and
MT-V potentials are spin-independent, while the ATS3 and Minnesota
potentials are spin-dependent. The parameters of the interactions are
tabulated in Table I. We choose $\hbar^2/m=41.47$ MeV fm$^2$. The Coulomb
interaction is included only in calculations with the Minnesota potential
where point charges are assumed and $e^2=1.44$ MeV fm.
\par\indent
The spins (and isospins) of the nucleons are coupled through successive
intermediate couplings. The spin couplings  up to $N\!\!=\!\!7$ nucleons are
tabulated in Table II. One naturally expects and test calculations show
that, without spin-isospin coupling, the energy convergence is much slower.
The number of spin-isospin configurations rapidly increases with $N$. In
the case
of $^6$He, for example, assuming $S=0$ and $T=1$, the wave function has
$5\!\times\! 9=45$ spin-isospin components. The number of components
becomes even
higher if the interaction has non-central spin-orbit, tensor, etc. parts.
The nonlinear parameters are not optimized with respect to spin-isospin
components, but rather, for each trial choice of the matrix $A$, we select
the spin-isospin component
that gives the lowest energy. Because the matrix elements between different
spin-isospin components differ only in linear factors ($c_{\kappa}$ in
Eq. (20)), the calculation
of the matrix elements of each spin-isospin component of the wave function
requires essentially the same computational effort as that needed by the
calculation for only one component.

\par\indent
Each calculation has been repeated several times starting
from different random points to check the energy convergence.
The energy as a function of the number of basis states
is shown in Fig. 2 for the case of $^6$Li with the Volkov potential.
The energies on different random paths, after a few initial steps,
approach to each other and converge to the final
solution. The energy difference between two random paths as well
as the tangent of the curves give us some information on the accuracy
of the method on a given size of the basis. The root mean square (rms)
radius of the few-nucleon system is calculated in each step and
found to be rapidly convergent to its final value. By increasing the
basis size the results can be arbitrarily improved when needed.
\par\indent
The number of basis states required to reach energy convergence increases
with the number of particles but it depends on the form of the interaction
as well. This latter property is illustrated in Fig. 3 for the case
of the $\alpha$-particle. The soft-core Volkov potential shows rapid
convergence, while the hard-core ATS3 interaction requires more basis states
to get an accurate solution. The relatively fast convergence for the MT-V
potential of a strong repulsion can be explained by the simplicity -- the
spin-independent nature -- of this interaction.
\par\indent
In the following we show tables for the ground-state energies $E$ and point
matter rms radii $\langle r^2\rangle^{1/2}$. The basis dimension ${\cal K}$
of the SVM listed in the tables is such that, beyond it, the energy and
the radius do not change in the digits shown. Table III shows our results
(SVM), together with results of others, for the
application of the spin-averaged MT-V potential  \cite{mt}  to $N$=2--7-nucleon
systems. For three-body systems, the solution of the Faddeev equation
is known to be the method of choice, but the SVM can
easily yield the same accurate energy. As the MT-V potential is a preferred
benchmark test of the few-body calculations, there are numerous solutions
available. Table III includes a few of the most accurate results.
  The nice agreement for four-nucleon case corroborates that the SVM is as
accurate as the direct solution of the FY equations
\cite{glocka}, the method of
the Amalgamation of Two-body correlations into Multiple Scattering (ATMS)
\cite{atms} process or the VMC \cite{wiringa} and GFMC \cite{gfma} method.
The basis used
in the Coupled-Rearrangement-Channel Gaussian-basis (CRCG) variational
method \cite{crcbgv} is similar to that of the SVM, but
the Gaussian parameters follow geometric progressions.
The fact that the basis size needed in the SVM is much smaller
proves the efficiency of the selection procedure.
The results of the VMC calculation for the five- and six-nucleon systems
are also in good agreement with the results of SVM. The MT-V potential
has no exchange term; therefore,
unlike the nature, it renders the five-nucleon system bound,
and the nucleus tends to collapse as the binding energy increases with the
number of particles.
\par\indent
The next example is the Volkov potential which, due to its very soft core,
is the most readily solvable case. This simple potential is widely used in
model calculations for light nuclei. As one sees in Table IV, the results
of SVM agree with those of other calculations, especially with the one
using hyperspherical harmonics (HH) functions. The number of basis states
needed to reach convergence is remarkably smaller than in the case of
the  MT-V potential. Without the Majorana exchange term
($M=0$) this potential also leads to a collapsing
system. By setting the Majorana parameter to $M$=0.6, a commonly used value
to get the correct binding energy, one may obtain more reasonable energies.
The Volkov potential with $M$=0.6 does not change the energies of $N$=2$-$4
nuclei, does not bind $^5$Li in accordance with the nature, but does bind
the $^6$Li ground state ($E$=$-$31.82 MeV, ${\langle r^2 \rangle}^{1/2}$
=2.69 fm).
\par\indent
Another potential that is often used in test calculation is the ATS3
potential. We have challenged the SVM to get solution for this case
because, unlike the Volkov, this spin-dependent potential has a relatively
strong repulsive core (see Table I). The solution, although on a somewhat
larger basis, can easily be obtained, and
it is in good agreement with those of other methods in the $N$=3 system
as shown in Table V. We note, however, that the energy of the SVM is
significantly lower than the ones of other methods for the
$\alpha$-particle. This may be due to the strong repulsion of this
potential. For example, the FY calculation \cite{glocka} agrees
with ours for the
MT-V potential, but shows a noticeable difference in the case of the
ATS3 potential. The variational calculation \cite{etbm} using a
correlation factor also misses considerable energy for the ATS3 potential.
Surprisingly, inspite of its exchange part, this potential also binds the
five-nucleon system and overbinds the six-nucleon systems very much.
\par\indent
The last example for the few-nucleon system uses the more realistic
Minnesota potential\cite{minesota}, which is a central interaction of
Gaussian form,
containing space-, spin-, and isospin-exchange operators (see Table I).
The Minnesota  potential has often been used in cluster-model
calculations of light nuclei. Table VI shows results with this potential,
where the Coulomb interaction between protons is also
included. All possible spin and isospin configurations
are allowed for and all spherical harmonics that give
non-negligible contribution are included in $\theta_{LM_L}({\bf x})$.
Since the method has proved to be accurate and reliable
for other potentials, it is justifiable to view
these results as testing the interaction rather than the method,
and hence the results are compared with experimental data.
The energy and the radius of triton and $\alpha$-particle converge, with
small
bases, to realistic values. The Minnesota potential, correctly, does not
bind the $N$=5 system, but it binds $^6$He and
slightly overbinds $^6$Li. The radius of $^6$He is found to be much larger
than that of $^4$He, consistently with the halo structure of $^6$He
\cite{SVAO}. It is for the first time that
the Minnesota force is tested without assuming any cluster structure or
restricting the model space by any other bias. The agreement is surprisingly
good not only with experiment but also with cluster-model calculations for all
nuclei \cite{LKBD}.
\par\indent
It is interesting to note that none of these simple potentials binds the
four-neutron system. The Volkov potential, for example, is so strong that it
binds the singlet
two-neutron system, but it does not allow the neutrons to form a
four-neutron bound state due to the Pauli principle.
\par\indent
As an example for bosonic nuclear few-body system, we consider the
case of structureless $\alpha$-particles interacting via the state-independent
potential;
\begin{equation}
V(r)=500\,\,{\rm exp}\,[-(0.7r)^2\,]-130\,\,{\rm exp}\,[-(0.475r)^2\,] \ \ \ \
({\rm MeV}),
\end{equation}
where $r$ is in fm. This potential is taken from Ali-Bodmer's $S$-state
potential\cite{AB}.
It has a repulsive core which is about 370 MeV high
and extends up to 2 fm. The repulsive core prevents the $\alpha$-particles from
collapsing.
The results are compared in Table VII. Our calculation agrees with the
ATMS result for the $N=$3 and 4 systems.

\subsection{Coulombic systems}
The results for the long range $1/r$ potential are collected in this
subsection. The first example is the polyelectric system $(me^+,ne^-)$.
The possibility that $m$ positrons and $n$ electrons form a bound system
was originally suggested by
Wheeler \cite{wheeler} and this question has been extensively
studied since then. Besides the trivial and analytically solvable  $m=1$
and $n=1$
case, the existence of the positronium negative ion ($m=1,n=2$) was also
predicted by Wheeler \cite{wheeler}.
Dozens of works have attempted to solve the $e^+,e^-,e^-$ Coulombic
three-body problem, continuously
refining the accuracy of the calculated binding energy \cite{epe,ho}.
Despite of numerous attempts, no one has obtained bound states for the
polyelectric system of more than four particles.
The positronium negative ion has experimentally been observed \cite{mills}.

\par\indent
The binding energy of the positronium molecule $(2e^+,2e^-)$ was first
calculated by Hylleraas and Ore \cite{hyll}. To date, the positronium
molecule has not been directly observed, and this fact intensifies the
theoretical interest to solve this Coulombic four-body problem.
This molecule is short-lived because the electron and positron may
annihilate.  Unlike the positronium ion, the positronium molecule is
neutral, and therefore the best
chance to distinguish it from the positronium itself is related to their
different lifetimes.
The QED formulae to determine the probability of a pair annihilation in
the positronium
molecule through a $k$-photon process ($k=0$,1,2,...) would require a highly
accurate wave function \cite{frolov95}.
\par\indent
In Table VIII we compare our results to the most precise calculations found in
the literature.
The correlated Gaussian function without the  polynomial part
($K=0$ in Eq. (6))
is known to poorly represent the Coulomb cusps \cite{KA91}. To improve the cusp
properties
the trial function with $K=0,1,2$ polynomials has been used.
\par\indent
As is shown in Table VIII, our calculation reproduces the first six digits
of the variational calculation of Ref.~\cite{epe} for the ground state of
(2e$^+$,e$^-$),  and the rms radius also agrees with it.
There are two recent variational calculations \cite{epep,KA93} for the
positronium molecule
using the correlated Gaussian functions. In these works the nonlinear
parameters were determined by optimatization. To compare our calculations
to theirs directly, the value of $K=0$ was chosen and the same basis size
(${\cal K}$=300) was used. Our result is slightly better than the energy
obtained by them
and this reinforces the reliability and powerfulness of the random selection of
the nonlinear
parameters.
The number of nonlinear parameters of this case is ${\cal K}\!\!\times
\!\!(4\!\!\times \!\!3)/2=1800$. The complete optimatization of the parameters
is, of course, superior to the SVM. Test calculations show that, provided
the number of parameters is low, that is, a
full optimatization is feasible, the optimatization finds lower energy.
But when the number of parameters becomes high, the full optimatization
becomes
less and less practical partly because it fails to find true minimum and
partly because the computation becomes too excessive.
\par\indent
We found no bound states for the ($3e^+,2e^-$) and ($3e^+,3e^-$)
systems. The energy of ($3e^+,3e^-$), for example,
converges to the sum of the energy of a dipositronium molecule and of
a positronium (0.515989 a.u.+0.25 a.u.=0.765989 a.u.). Allowing the selection
of the parameters from a larger region increases the rms radius, which is
typical of an unbound state. The system of
a negative and a positive positronium ion thus forms no bound state but
dissociates into a dipositronium molecule and a positronium.
\par\indent
Calculations for $L=1$ state also fails to find a bound system.
This result entails that the Coulomb force cannot bind more than four
particles out of identical charged fermions and their antiparticles.
\par\indent
To examine the role of the Pauli principle in preventing five-electron-positron
system from forming bound states, we repeated the same calculation
replacing the fermions by bosonic equivalents.
On a different scale, these systems may be identified, e.g., as the
systems of $\pi^-$ and $\pi^+$ with their strong interaction neglected
\cite{boson}. Such bosons turn out to form bound states even for $N=5$.
As is expected, the radius of the charged boson
system decreases as the number of particles increases. It is interesting
to note that the energies of bosonic and fermionic systems are
equal for $N=3$ and for $N=4$. The reason is that the energy minimum belongs
to the same spatial configurations, that is, to a triangular pyramid for
$N=4$, for example \cite{ho86}.
\par\indent
In Table IX we show results for bosonic and fermionic systems with a
purely attractive
$Gm^2/r$ (``gravitational") interaction.
An $N$-body system of identical particles bound together by attractive
pair potentials always collapses in large-$N$ limit
(the binding energy per particle rises with $N$ to infinity),
even if the particles are fermions. Self-gravitating boson systems have
recently attracted some interest \cite{selfgrav}. For these systems,
both variational lower and upper bounds are available.
In this case even the five-fermion system
is bound. Thus the lack of bound states in
five-electron-positron systems is a joint effect of the antisymmetry
and of the repulsion between identical particles.
\par\indent
Finally, we mention an example involving an excited state. With ${\cal K}=500$,
the SVM gives the energies of the ground and first excited states of the
$t+d+\mu^-$ system as $-$111.3640 and $-$100.9121 a.u., which are respectively
compared to $-$111.364342 and $-$100.916421 a.u. of the CRCG result
\cite{dtmu} with ${\cal K}=1442$, while the configuration-space Faddeev
calculation \cite{faddeev}  gives $-$111.36 a.u. for the ground state but no
information for the excited state.

\section{Summary}
\par\indent
We have formulated a variational calculation for few-body systems using the
stochastic variational method on the
correlated Gaussian basis. We have  demonstrated the versatility of the
correlated Gaussians and the efficiency of the stochastic variation by
various numerical examples for $N=2-7$-particle systems. All the details of
both formulation and calculational procedure are included to make this
paper self-contained and easily reproducible.
\par\indent
The comparison with other calculations has corroborated the accuracy and
efficiency of the method. In none of the test cases has the present method
proved to be inferior to any of the alternative methods, and yet the method
does not require excessive computational effort.
\par\indent
The correlation between the particles plays an important role in describing
the few-body system realistically. It has been taken into account in the
framework of the correlated Gaussian functions. The correlated Gaussians
are constructed from
products of the Gaussian wave-packet single-particle functions through an
integral transformation, which has enabled us to evaluate the
center-of-mass motion free matrix elements analytically starting from the
single-particle level. The nonlinear parameters of the correlated Gaussians
have been selected by the stochastic variational method with a trial and
error procedure. The success of the method using the correlated Gaussian
basis is probably due to the fact that none of the Gaussians is
indispensable, that is,
there are different sets of the Gaussian parameters that represent the wave
function equally well.

\par\indent
The method presented in this paper can be useful to solve few-body problems
in diverse fields of physics such as description of microclusters,
non-relativistic quark model, and halo nuclei. Among others, the most
important application is the
solution of the nuclear few-body problem, that is, a description of light
nuclei by using ${\it realistic}$ nucleon-nucleon potentials. In this case
one has to take into account both short-range repulsion and higher orbital
angular momenta required by the non-central components. Our test examples
show that the correlated Gaussian basis function might be a suitable
candidate to cope with these requirements. As is explained
in Appendix B, the evaluation of the matrix elements for the non-central
potentials poses no serious problem, and calculations including such
potentials for nuclear few-body systems are under way.
\par\indent
The limitations of
the present method are those implied by the basis size and by the computer
memory to store the matrix elements in the generator coordinate space. The
limitations may become excessive as the number of particles and/or spin
and isospin configurations become large.
\par\indent
To extend the method to nuclei of larger mass number in an approximate
way, one can
freeze part of the model space for a group of nucleons (cluster). One can
omit, for example, some of the spin or isospin channels. It might be a
good approximation to consider only those spin channels where the spins of
the like nucleon pairs are coupled to zero. One can also restrict the intrinsic
spatial motion of a cluster by fixing the nonlinear parameters to some
appropriate values. One can introduce $N$ clusters and place the nucleons
of each cluster into a common harmonic oscillator well, for example.  The
microscopic multicluster model is based on this approximation. The matrix
elements needed in this multicluster model are given as a special case of those
presented
in this paper. In fact, one only needs to choose the single-particle
generator coordinates, $({\bf s}_1,...,{\bf s}_A)$, such that
\begin{equation}
{\bf s}_{n_{i-1}+1}={\bf s}_{n_{i-1}+2}=\cdot\cdot\cdot=
{\bf s}_{n_{i-1}+n_i} \ \ \ \ (i=1,...,N),
\end{equation}
where $n_i$ is the number of nucleons in
the ${\it i}$th cluster ($n_0=0,\ \  n_1+\cdot\cdot\cdot+n_N=A$), and needs
to couple appropriately the spins and isospins of the nucleons in each
cluster to the spin and isospin of the cluster. The microscopic
multicluster model has been successfully applied for description of the
structure of light nuclei (see, for example, \cite{SVAO,Brink,wild}).
\par\indent
\par\indent
Finally, we summarize some merits of our method in the following:
\begin{description}
\item{(i)} Fully analytical calculational scheme; this plays a major
role in the high speed and accuracy of the calculation.
\item{(ii)} Universality of the scheme. One needs to introduce
no change, for example, between describing a multinucleon system
and a Coulombic few-body system. It is easily adaptable to identical
or non-identical particles, to fermions or bosons or mixed systems.
The masses of particles may be different, yet no problem with
the center-of-mass motion arises.
\item{(iii)} No expansion of the interaction is needed, and thus
no problems in partial-wave truncation arise.
\item{(iv)} The convergence of the energy is fast. If one needs just a
2--3-digit estimate of the energy, it is enough to use a very small basis.
\item{(v)} The method is also accurate for excited states, which are
obtained simultaneously with the same diagonalization (provided their
angular momenta and parities are the same as those of the ground state;
but only such excited states may be problematic).
\item{(vi)} The wave function is obtained in a compact, analytical form.
It is then easy to use it in calculations of physical properties.
It is ``portable'', reproducible, and easily testable.
\end{description}

\section*{Appendix A. Single-particle matrix elements}
The aim of this appendix is to list the single-particle matrix elements
between Gaussian wave packets (Eq. (17)).
The Gaussian packets are generalized in this appendix to have different width
parameters and the expressions are therefore slightly more general
than needed
in the formulae of the main text. These single-particle matrix elements are,
however, required for treating particles of unequal masses, in which
the width parameter $\nu$ belonging to the particle of mass $m$ is
to be chosen by Eq. (59). The overlap of two Gaussian wave packets is
\begin{equation}
\langle \varphi_{{\bf s}_1}^{\nu_1} \vert \varphi_{{\bf s}_2}^{\nu_2} \rangle
=\left({2 \sqrt{\nu_1 \nu_2} \over \nu_1 +\nu_2}\right)^{3/2}
 {\rm exp}\,\left(-{\nu_1 \nu_2 \over \nu_1 +\nu_2} ({{\bf s}_1}-
{{\bf s}_2})^2\,\right) .
\end{equation}
The matrix element of the kinetic energy operator $(T=-\frac{{\hbar}^2}
{2M}\Delta)$ reads as
\begin{equation}
\langle \varphi_{{\bf s}_1}^{\nu_1} \vert T\vert\varphi_{{\bf s}_2}^{\nu_2}
\rangle
={\hbar^2  \over 2M }
{2\nu_1\nu_2 \over \nu_1 +\nu_2}
\left( 3-{2 \nu_1 \nu_2\over \nu_1+\nu_2} ({\bf s}_1-{\bf s}_2)^2 \right)
\langle \varphi_{{\bf s}_1}^{\nu_1} \vert
\varphi_{{\bf s}_2}^{\nu_2} \rangle .
\end{equation}
The matrix element of the square radius becomes
\begin{equation}
\langle \varphi_{{\bf s}_1}^{\nu_1} \vert {\bf r}_1^2 \vert
\varphi_{{\bf s}_2}^{\nu_2} \rangle
= {1 \over 2(\nu_1 +\nu_2)}
\left( 3+{2 \over \nu_1+\nu_2} (\nu_1{\bf s}_1+\nu_2{\bf s}_2)^2 \right)
\langle \varphi_{{\bf s}_1}^{\nu_1} \vert
\varphi_{{\bf s}_2}^{\nu_2} \rangle .
\end{equation}
The two-particle matrix element of a $\delta$-function  is given by
\begin{eqnarray}
& &\langle \varphi_{{\bf s}_1}^{\nu_1}\varphi_{{\bf s}_2}^{\nu_2}
\vert \delta({\bf r}_1-{\bf r}_2-{\bf r})\vert
\varphi_{{\bf s}_3}^{\nu_3}\varphi_{{\bf s}_4}^{\nu_4} \rangle
=
\left({(\nu_1+\nu_3)(\nu_2+\nu_4) \over
(\nu_1+\nu_2+\nu_3+\nu_4) \pi }\right)^{3/2} \nonumber \\
&\times&
{\rm exp}\,\Biggl(-{(\nu_1+\nu_3)(\nu_2+\nu_4) \over
\nu_1+\nu_2+\nu_3+\nu_4}
\Bigl({\bf r}-{\nu_1{\bf s}_1+\nu_3{\bf s}_3\over \nu_1+\nu_3}
+{\nu_2{\bf s}_2+\nu_4{\bf s}_4\over \nu_2+\nu_4}\Bigr)^2\,\Biggr)
\langle \varphi_{{\bf s}_1}^{\nu_1} \vert \varphi_{{\bf s}_3}^{\nu_3}
\rangle
\langle \varphi_{{\bf s}_2}^{\nu_2} \vert \varphi_{{\bf s}_4}^{\nu_4}
\rangle .
\end{eqnarray}

\section*{Appendix B. Matrix elements of the two-body potentials}
The scope of this appendix is the calculation of the matrix elements
of the different ingredients of the  two-nucleon interaction
between Gaussian wave packets. Most of the widely used
coordinate-space two-nucleon interactions consist of central- ($O^c_{12}$),
tensor- ($O^t_{12}$),
spin-orbit- ($O^b_{12}$), ${\bf L}^2$- (or ${\bf p}^2$) ($O^q_{12}$) and
quadratic spin-orbit  ($O^{bb}_{12}$) type potentials.
(We follow the abbreviated notations
-- $c,t,b,q,bb$ -- invented by Urbana-Argonne group \cite{LP81,WSA84}. The
definition of these operators is given in Table X.)
These potential terms might be multiplied by the
\mbox{\boldmath{$\tau$}}$_1\cdot$\mbox{\boldmath{$\tau$}}$_2$ ($O^\tau_{12}$),
\mbox{\boldmath{$\sigma$}}$_1\cdot$\mbox{\boldmath{$\sigma$}}$_2$
($O^\sigma_{12}$),
or \mbox{\boldmath{$\tau$}}$_1\cdot$\mbox{\boldmath{$\tau$}}$_2$
\mbox{\boldmath{$\sigma$}}$_1\cdot$\mbox{\boldmath{$\sigma$}}$_2$
($O^{\tau \sigma}_{12}$) spin-isospin operators
and then one ends up with the general form
\begin{equation}
V_{ij}=
\sum_{p} \int d{\bf r} V^{p}(r) \delta({\bf r}_i-{\bf r}_j-{\bf r})
O_{ij}^{p} ,
\end{equation}
where $p=c,c\tau,c\sigma,c\tau\sigma,t,t\tau,b,b\tau,q,q\tau,q\sigma,
q\tau\sigma,
bb,bb\tau$ is the short-hand notation to specify the component and $V^p(r)$
is the corresponding radial form factor. By using Eqs. (16), (17), (68),
(69), and after some straightforward transformation, the matrix element of
this interaction   can be written as
\begin{eqnarray}
& &\langle
{\hat \phi}^{\nu}_{{\bf s}_1\sigma_1\tau_1}
{\hat \phi}^{\nu}_{{\bf s}_2\sigma_2\tau_2}
\vert
V_{12}
\vert
{\hat \phi}^{\nu}_{{{\bf s}_1}'\sigma_1'\tau_1'}
{\hat \phi}^{\nu}_{{{\bf s}_2}'\sigma_2'\tau_2'}
\rangle =
\sum_{p} \int d{\bf r} V^{p}(r) f_{p}(r){\cal M}({\bf r})\nonumber \\
&\times&\langle\chi_{{1\over 2}\sigma_1}{\cal X}_{{1\over 2}\tau_1}\,
\chi_{{1\over 2}\sigma_2}{\cal X}_{{1\over 2}\tau_2}\vert\,
r^{2}B^{p}+\sum_{l=0}^{2}\sum_{m=-l}^{l}
(-1)^{m}r^{l}Y_{lm}({\hat {\bf r}}) C^{p}_{l\,-m}\,\vert\,
\chi_{{1\over 2}\sigma'_1}{\cal X}_{{1\over 2}\tau'_1}\,
\chi_{{1\over 2}\sigma'_2}{\cal X}_{{1\over 2}\tau'_2}\rangle,
\end{eqnarray}
with
\begin{equation}
{\cal M}({\bf r})=\Bigl({\nu \over \pi}\Bigr)^{3/2}
{\rm exp}\,\Biggl(-\nu r^2+\nu ({\bf s}_1\!\!-\!\!{\bf s}_2\!\!+\!\!
{\bf s}'_1\!\!-\!\!{\bf s}'_2)\cdot{\bf r}-{\nu \over 4}
({\bf s}_1\!\!-\!\!{\bf s}_2\!\!+\!\!{\bf s}'_1\!\!-\!\!{\bf s}'_2)^2\,
\Biggr)\,\langle \varphi_{{\bf s}_1}^{\nu} \vert \varphi_{{\bf s}'_1}
^{\nu} \rangle
\langle \varphi_{{\bf s}_2}^{\nu} \vert \varphi_{{\bf s}'_2}^{\nu}
\rangle .
\end{equation}

The  $B^{p}$ and $C^{p}_{lm}$, independent of ${\bf r}$, are the operators
in spin-isospin space and are listed in Table X  for the most important terms
($c,t,b,q,bb$). The remaining terms
can be easily derived by multiplying these operators by the appropriate
$\mbox{\boldmath $\sigma$}$ and $\mbox{\boldmath $\tau$}$ operators. The
function $f_p(r)$  has the simple form:
$f_p(r)=r^{-2}$ if $p=t$ or $t\tau$, and $f_p(r)=1$ otherwise.
\par\indent
The calculation of matrix elements of the operators
appearing in Table X in the spin part is easily done with the
Clebsch-Gordan coefficient. They are given below by suppressing the spin
function $\chi_{\frac{1}{2} \sigma_i}\,(m_i=\frac{1}{2} \sigma_i)$.

\begin{eqnarray}
& &\langle m_1\,m_2\,\vert\,
{\mbox{\boldmath $\sigma$}_1}\cdot{\mbox{\boldmath $\sigma$}_2}\,
\vert \,m'_1\,m'_2\,\rangle \nonumber \\
&=&3(-1)^{m_1-m'_1}\delta_{m_1+m_2,m'_1+m'_2}
\langle {1 \over 2}\,m'_1\,1\,m_1\!\!-\!\!m'_1\,\vert\,
{1 \over 2}\,m_1\rangle
\langle {1 \over 2}\,m'_2\,1\,m_2\!\!-\!\!m'_2\,\vert\,
{1 \over 2}\,m_2\rangle.
\end{eqnarray}

\begin{eqnarray}
& &\langle m_1\,m_2\,\vert\,
\left[{\mbox{\boldmath $\sigma$}_1}\times{\mbox{\boldmath $\sigma$}_2}
\right]^{(2)}_{m}\, \vert \,m'_1\,m'_2\,\rangle=3\,\delta_{m,m_1+m_2-m'_1-m'_2}
\nonumber \\
&\times&\langle {1 \over 2}\,m'_1\,1\,m_1\!\!-\!\!m'_1\,\vert\,
{1 \over 2}\,m_1\rangle
\langle {1 \over 2}\,m'_2\,1\,m_2\!\!-\!\!m'_2\,\vert\,
{1 \over 2}\,m_2\rangle
\langle 1\,m_1\!\!-\!\!m'_1\,1\,m_2\!\!-\!\!m'_2\,\vert\,2\,m\rangle.
\end{eqnarray}

\begin{eqnarray}
& &\langle m_1\,m_2\,\vert\,
\left[{\bf x}\times
({\mbox{\boldmath $\sigma$}_1}+{\mbox{\boldmath $\sigma$}_2})
\right]^{(1)}_{m}\, \vert\,m'_1\,m'_2\,\rangle=\sqrt{3}\, \sum_{q_1,q_2=-1}
^{1}({\bf x})_{q_1} \nonumber \\
&\times&\langle 1\,q_1\,1\,q_2\,\vert\,1\,m\rangle
\Bigl\{\delta_{m_2,m'_2}
\langle{1\over 2}\,m'_1\,1\,q_2\,\vert\,{1\over 2}\,m_1\rangle+
\delta_{m_1,m'_1}
\langle{1\over 2}\,m'_2\,1\,q_2\,\vert\,{1\over 2}\,m_2\rangle\Bigr\}.
\end{eqnarray}

\begin{eqnarray}
& &\langle m_1\,m_2\,\vert\,
({\bf x}\cdot{\mbox{\boldmath $\sigma$}_2})
{\mbox{\boldmath $\sigma$}_{1_{m}}}+
({\bf x}\cdot{\mbox{\boldmath $\sigma$}_1})
{\mbox{\boldmath $\sigma$}_{2_{m}}}
\, \vert \,m'_1\,m'_2\,\rangle \nonumber \\
&=&3\,\Bigl\{(-1)^{m_2-m'_2}\,\delta_{m,m_1-m'_1}({\bf x})_{m'_2-m_2}+
(-1)^{m_1-m'_1}\,\delta_{m,m_2-m'_2}({\bf x})_{m'_1-m_1}\Bigr\} \\
&\times&\langle {1 \over 2}\,m'_1\,1\,m_1\!\!-\!\!m'_1\,\vert\,
{1 \over 2}\,m_1\rangle
\langle {1 \over 2}\,m'_2\,1\,m_2\!\!-\!\!m'_2\,\vert\,
{1 \over 2}\,m_2\rangle . \nonumber
\end{eqnarray}

\begin{eqnarray}
& &\langle m_1\,m_2\,\vert\,
({\bf x}\cdot{\mbox{\boldmath $\sigma$}_1})
({\bf x}\cdot{\mbox{\boldmath $\sigma$}_2})
\, \vert \,m'_1\,m'_2 \rangle \,
=3\,(-1)^{m_1-m'_1+m_2-m'_2} \nonumber \\
&\times&({\bf x})_{m'_1-m_1}({\bf x})_{m'_2-m_2}
\langle {1 \over 2}\,m'_1\,1\,m_1\!\!-\!\!m'_1\,\vert\,
{1 \over 2}\,m_1\rangle
\langle {1 \over 2}\,m'_2\,1\,m_2\!\!-\!\!m'_2\,\vert\,
{1 \over 2}\,m_2\rangle .
\end{eqnarray}

\begin{eqnarray}
& &\langle m_1\,m_2\,\vert\,
\left[\,\left[{\bf x}\times{\mbox{\boldmath $\sigma$}_1}\right]^{(1)}
\times\left[{\bf x}\times{\mbox{\boldmath $\sigma$}_2}\right]^{(1)}
\,\right]^{(2)}_{m}
\, \vert \,m'_1\,m'_2\,\rangle  \nonumber \\
&=&3\,\langle {1 \over 2}\,m'_1\,1\,m_1\!\!-\!\!m'_1\,\vert\,
{1 \over 2}\,m_1\rangle
\langle {1 \over 2}\,m'_2\,1\,m_2\!\!-\!\!m'_2\,\vert\,
{1 \over 2}\,m_2\rangle\,\nonumber \\
&\times& \sum_{q_1,q_2=-1}^{1}({\bf x})_{q_1}({\bf x})_{q_2}
\langle 1\,q_1\!\!+\!\!m_1\!\!-\!\!m'_1\,1\,q_2\!\!+\!\!m_2\!\!-\!\!
m'_2\,\vert\,2\,m\rangle \\
&\times&\langle
1\,q_1\,1\,m_1\!\!-\!\!m'_1\,\vert\,1\,q_1\!\!+\!\!m_1\!\!-\!\!m'_1
\rangle
\langle 1\,q_2\,1\,m_2\!\!-\!\!m'_2\,\vert\,1\,q_2\!\!+\!\!m_2\!\!-\!\!m'_2
\rangle. \nonumber
\end{eqnarray}
Here ${\bf x}$ is a 3-dimensional vector and $({\bf x})_m$ stands for its
spherical component $\sqrt{4\pi\over 3} x Y_{1m}({\hat{\bf x}})$.

\section*{Appendix C. Matrix elements of Slater determinants}
In this appendix we briefly outline the calculation of the matrix elements
between Slater determinants of Eq. (18) and show their concrete functional
form, that is,
the dependence on the generator coordinates ${\bf s}$. We assume that the
width parameter of the Gaussian wave packet is chosen to be the same for
all nucleons.
The overlap of two Slater determinants is equal to the determinant of
the matrix of the single-particle overlaps:
\begin{equation}
\langle \phi_{\kappa} ({\bf s}_1,...,{\bf s}_N) \vert
\phi_{\kappa'}({{\bf s}'}_1,...,{{\bf s}'_N}) \rangle
={\rm det} \lbrace B \rbrace,
\end{equation}
where
\begin{equation}
B_{ij}=
\langle {\hat\varphi}_{{\bf s}_i\sigma_i\tau_i}^{\nu}
\vert
{\hat\varphi}_{{{\bf s}'}_j{\sigma}'_j{\tau}'_j}^{\nu} \rangle
 \ \ \ \ \ \ \ (i,j=1,...,N).
\end{equation}
By using the definition of the determinant, this can be rewritten as
\begin{equation}
\langle \phi_{\kappa} ({\bf s}_1,...,{\bf s}_N) \vert
\phi_{\kappa'}({{\bf s}'}_1,...,{{\bf s}'_N}) \rangle
= \sum_{P}^{N!} {\rm sign}(P)
\langle {\hat \varphi}_{{\bf s}_1\sigma_1\tau_1}^{\nu} \vert
{\hat\varphi}_{{{\bf s}'}_{p_{1}}{\sigma}'_{p_1}{\tau}'_{p_1}}^{\nu}
\rangle...
\langle {\hat\varphi}_{{\bf s}_N\sigma_N\tau_N}^{\nu} \vert
{\hat\varphi}_{{{\bf s}'}_{p_{N}}{\sigma}'_{p_{N}}{\tau}'_{p_{N}}}
^{\nu} \rangle ,
\end{equation}
where $(p_{1}\cdot \cdot \cdot p_{N})$ is the permutation $P$ of the set
$(1\cdot \cdot \cdot N)$.
Substituting the overlap of the single-particle overlaps of Eq. (65)
into Eq. (80) yields an explicit formula for the overlap of the
Slater determinants:
\begin{equation}
\langle \phi_{\kappa} ({\bf s}_1,...,{\bf s}_N) \vert
\phi_{\kappa'}({{\bf s}'}_1,...,{{\bf s}'_N}) \rangle =
\sum_{P}^{N!} C_{P}
{\rm e}^{-{1\over 2} \tilde{\bf s} A_{P} {\bf s}} ,
\end{equation}
where the matrix $A_P$ is defined by
\begin{equation}
(A_P)_{ij}=(A_P)_{N+i,N+j}= \nu \delta_{ij},\ \ \ \
(A_P)_{i,N+j}=(A_P)_{N+j,i}=-\nu \delta_{j,p_{i}}
\ \ \ \    (1\le i,j \le N),
\end{equation}
and
\begin{equation}
C_{P}={\rm sign}(P)
\delta_{{\sigma_1}{\sigma'}_{p_1}} \delta_{{\tau_1}{\tau'}_{p_1}}
\cdot\cdot\cdot \delta_{{\sigma_N}{\sigma'}_{p_N}}
\delta_{{\tau_N}{\tau'}_{p_N}}.
\end{equation}
The orthogonality of the spin-isospin functions greatly reduces the number
of terms in the summation over $P$.
\par\indent
The matrix element of the kinetic energy operator is
\begin{equation}
\sum_{i=1}^{N}
\langle \phi_{\kappa} ({\bf s}_1,...,{\bf s}_N) \vert T_i \vert
\phi_{\kappa'}({{\bf s}'}_1,...,{{\bf s}'_N}) \rangle =
\sum_{i=1}^{N}
\sum_{j=1}^{N}
\langle{\hat \varphi}_{{{\bf s}_i}{\sigma_i}{\tau_i}}^{\nu}
\vert T \vert {\hat \varphi}_{{{{\bf s}'}_j}{\sigma'_j}
{\tau'_j}}^{\nu} \rangle
(-1)^{i+j} {\det} \lbrace B^{ij}\rbrace,
\end{equation}
where $B^{ij}$ is obtained by omitting the $i$th row and the $j$th column
of the matrix $B$ defined in Eq. (79).
The substitution of the single-particle overlaps and the
single-particle matrix elements of the kinetic energy operator enables us
to obtain
\begin{equation}
\sum_{i=1}^{N}
\langle \phi_{\kappa} ({\bf s}_1,...,{\bf s}_N) \vert T_i \vert
\phi_{\kappa'}({{\bf s}'}_1,...,{{\bf s}'_N}) \rangle =
{\hbar \omega \over 2}
\sum_{P}^{N!} C_{P} \left({3 \over 2}N -{1\over 2} \tilde{\bf s} A_{P}
 {\bf s}\right){\rm e}^{-{1\over 2} \tilde{\bf s} A_{P} {\bf s}}.
\end{equation}
Note that the coefficients $C_{P}$ and the matrices $A_P$ in Eqs. (81) and
(85) are the same. The manipulation similar to this leads us to the
matrix element of the square radius as
\begin{equation}
\sum_{i=1}^{N}
\langle \phi_{\kappa} ({\bf s}_1,...,{\bf s}_N) \vert {\bf r}_i^2 \vert
\phi_{\kappa'}({{\bf s}'}_1,...,{{\bf s}'_N}) \rangle =
{1 \over 2\nu}
\sum_{P}^{N!} C_{P} \left({3 \over 2}N -{1\over 2} \tilde{\bf s} A_{P}
 {\bf s} + \nu\tilde{\bf s}{\bf s} \right){\rm e}^{-{1\over 2} \tilde{\bf s}
A_{P} {\bf s}}.
\end{equation}
\par\indent
As is explained in Appendix B, the matrix element of any two-body
interaction between the Slater determinants may be reduced to the
following:
\begin{equation}
\sum_{i<j}^{N}
\langle \phi_{\kappa} ({\bf s}_1,...,{\bf s}_N)
\vert
\delta({\bf r}_i-{\bf r}_j-{\bf r})O_{ij}^p
\vert
\phi_{\kappa'}({{\bf s}'}_1,...,{{\bf s}'_N}) \rangle.
\end{equation}
The matrix element of Eq. (87) is called the correlation function of type
$p$ evaluated between the Slater determinants.  The calculation of this matrix
element can be done with the use of the basic
two-body matrix elements of Appendix B and the single-particle overlaps. We
will show, as an example, the case of $O_{ij}^p = 1$, i.e., Wigner-type
$\delta$-function two-body interaction $(p=c)$. Then we have
\begin{eqnarray}
& &\sum_{i<j}^{N}
\langle \phi_{\kappa} ({\bf s}_1,...,{\bf s}_N)
\vert
\delta({\bf r}_i-{\bf r}_j-{\bf r})
\vert
\phi_{\kappa'}({{\bf s}'}_1,...,{{\bf s}'_N}) \rangle
\nonumber \\
&=&\sum_{i<j}^{N} \sum_{k,l=1}^{N}
\langle{\hat \varphi}_{{{\bf s}_i}{\sigma_i}{\tau_i}}^{\nu}
{\hat\varphi}_{{{\bf s}_j}{\sigma_j}{\tau_j}}^{\nu}
\vert
\delta({\bf r}_i-{\bf r}_j-{\bf r})
\vert
{\hat\varphi}_{{{{\bf s}'}_k}{\sigma'_k}{\tau'_k}}^{\nu}
{\hat\varphi}_{{{{\bf s}'}_l}{\sigma'_l}{\tau'_l}}^{\nu} \rangle
(-1)^{i+j+k+l} {\det} \lbrace B^{ijkl}\rbrace,
\end{eqnarray}
where $B^{ijkl}$ is obtained by omitting the $i$th  and $j$th
rows and the $k$th
and $l$th columns of the matrix $B$. By substituting the explicit formula
of the ingredients, we arrive at the same form as given in Eq. (25):
\begin{eqnarray}
& &\sum_{i<j}^{N}
\langle \phi_{\kappa} ({\bf s}_1,...,{\bf s}_N)
\vert
\delta({\bf r}_i-{\bf r}_j-{\bf r})
\vert
\phi_{\kappa'}({{\bf s}'}_1,...,{{\bf s}'_N}) \rangle
\nonumber \\
&=&\bigl(\frac{\nu}{\pi}\bigr)^{3/2}{\rm e}^{-\nu r^2}
\sum_{P}^{N!}C_P\sum_{i<j}^{N}{\rm e}^{-{1\over 2} \tilde{\bf s}( A_{P}
 + B_{P}^{(ij)}) {\bf s}+ {\bf d}_{P}^{(ij)}\cdot {\bf r}},
\end{eqnarray}
with
\begin{equation}
{\bf d}_{P}^{(ij)}=\nu({\bf s}_i-{\bf s}_j + {\bf s}_{N+p_i}-{\bf s}_{N+p_j}).
\end{equation}
The coefficients $C_P$ and the matrices $A_P$ are the same as those that
appear in Eq. (81). The matrix $B_P^{(ij)}$ is defined by
\begin{eqnarray}
{(B_P^{(ij)})}_{kl}={(B_P^{(ij)})}_{lk}=\left\{
\begin{array}{ll}
\frac{\nu}{2} &\ \ \  (kl)=(ii), (jj), (N+p_i,N+p_i), (N+p_j,N+p_j),\\
        &  \ \ \ \ \ \  \  \ \ \ \ \   (i,N+p_i), (j,N+p_j).   \\
-\frac{\nu}{2} & \ \ \ (kl)=(ij), (i,N+p_j), (j,N+p_i), (N+p_i,N+p_j).  \\
0  & \ \ \   {\rm otherwise}.  \\
\end{array}
\right.
\end{eqnarray}

\section*{Appendix D. Evaluation of the overlap of the correlated Gaussians}
In this appendix we derive the overlap of the basis functions,
defined in Eqs. (2) and (6), to illustrate the calculation of the
matrix elements through the operations prescribed in Eq. (57).
For the sake of simplicity we work out the case
of overlap only, but the matrix elements of the kinetic and the
potential energy are not much more involved either and these
matrix elements can be calculated by repeating the steps
detailed here.
\par\indent
In Eq. (57) the matrix element of an operator ${\cal O}$ between the
correlated Gaussians are derived from the matrix elements
between the Gaussian packets by applying suitably chosen operations.
In the case of ${\cal O}=1$, upon substituting
the matrix elements in the generator coordinate space (Eq. (41)), Eq. (57)
reads as
\begin{eqnarray}
& &\langle {\cal A} \{f_{KLM}({\bf u},{\bf x},A){\chi}_{SM_S}{\cal X}_{TM_T}\}
\vert {\cal A'} \{f_{K'L'M'}({\bf u}',{\bf x},A')
{\chi}_{S'M'_S}{\cal X}_{T'M'_T}\}\rangle
\nonumber \\
&\sim &  \int\!\int d{\hat {\bf t}}d{\hat {\bf t}}'
Y_{LM}({\hat {\bf t}})^{*} Y_{L'M'}({\hat {\bf t}}') \\
&\times& \left( {d^{\kappa+\kappa'} \over d\alpha^{\kappa}d\alpha'^{\kappa'}}
{\rm e}^{-{1\over 2} \tilde{\bf T} C {\bf T}}
\,\int d{{\bf S}}
{\rm e}^{-{1\over 2} \tilde{{\bf S}} Q {{\bf S}}+\tilde{\bf T}{{\bf S}}}
\sum_{i=1}^{n_o}C_i^{(o)} {\rm e}^{-{1\over 2}\tilde{\bf S} B_i^{(o)}{\bf S}}
\right)_{\alpha=\alpha'=0 \atop t=t'=1},
\nonumber
\end{eqnarray}
with
\begin{equation}
\kappa=2K+L, \ \ \ \ \kappa'=2K'+L',
\end{equation}
where a constant factor is omitted for a moment.
By integrating over the generator coordinates ${\bf S}$ with the use of
Eq. (52), one readily obtains
\begin{eqnarray}
& &\langle {\cal A} \{f_{KLM}({\bf u},{\bf x},A){\chi}_{SM_S}{\cal X}_{TM_T}\}
\vert {\cal A'} \{f_{K'L'M'}({\bf u}',{\bf x},A')
{\chi}_{S'M'_S}{\cal X}_{T'M'_T}\}\rangle
\nonumber \\
&\sim & \int\!\int d{\hat {\bf t}}d{\hat {\bf t}}'
Y_{LM}({\hat {\bf t}})^{*} Y_{L'M'}({\hat {\bf t}}') \\
&\times& \left( {d^{\kappa+\kappa'} \over d\alpha^{\kappa}d\alpha'^{\kappa'}}
\,\sum_{i=1}^{n_o}C_i^{(o)}
\left({(2\pi)^{(2N-2)} \over {\rm det}(B_i^{(o)}+Q)}\right)^{3/2}
{\rm e}^{p_i\alpha^2+{{p_i}'}\alpha'^2+q_i\alpha\alpha' {\bf t}\cdot{\bf t'}}
\right)_{\alpha=\alpha'=0 \atop t=t'=1},\nonumber
\end{eqnarray}
where, by using Eqs. (54) and (55), we introduced the abbreviations
\begin{eqnarray}
& &p_i={1\over 2}\sum_{j,k=1}^{N-1} w_j  ((B_i^{(o)}+Q)^{-1}-C)_{jk} w_k,
\ \ \ {p_i}'={1\over 2}\sum_{j,k=N}^{2N-2} {w_j}'
((B_i^{(o)}+Q)^{-1}-C)_{jk} {w_k}' ,\nonumber \\
& &q_i=\sum_{j=1}^{N-1} \sum_{k=N}^{2N-2} w_j ((B_i^{(o)}+Q)^{-1}-C)_{jk}
 {w_k}' ,
\end{eqnarray}
with
\begin{equation}
w_k=\sum_{j=1}^{N-1}(\Gamma C)^{-1}_{kj}u_j, \ \ \ \
{w_k}'=\sum_{j=N}^{2N-2}(\Gamma C')^{-1}_{kj}{u_j}' ,
\end{equation}
in order to emphasize the $\alpha$- and $\alpha'$-dependence of the
resulting expression.
The differentiation with respect to $\alpha$ and $\alpha'$ can be promptly
given by expanding the exponential functions into power series:
\begin{eqnarray}
& &{d^{\kappa+\kappa'} \over d\alpha^{\kappa}d{\alpha'}^{\kappa'}}
{\rm e}^{p_i\alpha^2+{{p_i}'}\alpha'^2+q_i\alpha\alpha'{\bf t}\cdot{\bf t'}}
\nonumber \\
& = & \sum_{j,j',k}^{\infty}
{ p_i^j{{p_i}'}^{j'}({q_i}{\bf t}\cdot{{\bf t}'})^k \over j!{j'}!k!}
{(2j+k)!\over (2j+k-\kappa)!}
{(2j'+k)!\over (2j'+k-\kappa')!}
\alpha^{2j+k-\kappa} {\alpha'}^{2j'+k-\kappa'},
\end{eqnarray}
and therefore, after putting $\alpha=\alpha'=0$, the calculation of
Eq. (94) can be continued as
\begin{eqnarray}
& &\langle {\cal A} \{f_{KLM}({\bf u},{\bf x},A){\chi}_{SM_S}{\cal X}_{TM_T}\}
\vert {\cal A'} \{f_{K'L'M'}({\bf u}',{\bf x},A')
{\chi}_{S'M'_S}{\cal X}_{T'M'_T}\}\rangle
\nonumber \\
&\sim & \int\!\int d{\hat {\bf t}}d{\hat {\bf t}}'
Y_{LM}({\hat {\bf t}})^{*} Y_{L'M'}({\hat {\bf t}}') \\
&\times& \left(\sum_{i=1}^{n_o}C_i^{(o)}
\left({(2\pi)^{2N-2} \over {\rm det}(B_i^{(o)}+Q)}\right)^{3/2}
\kappa!\kappa'! \sum_{k}
{{p_i^{\kappa-k\over 2}{{p_i}'}^{\kappa'-k\over 2}}{q_i}^k \over
({\kappa-k \over 2})!
({\kappa'-k\over 2})! k!}
\left({\bf t}\cdot{\bf t}'\right)^{k}_{t=t'=1}\right), \nonumber
\end{eqnarray}
where the summation over $k$ runs from $0$ to max$(\kappa,\kappa')$
for those values that fulfill $(-1)^{k+\kappa}=1$ and $(-1)^{k+\kappa'}=1$.
\par\indent
The last step, the integration over the angles of ${\bf t}$ and ${\bf t}'$,
can be accomplished by applying Eq. (49) to express the scalar product
${\bf t}\cdot {\bf t}'$, and then we get the final expression:
\begin{eqnarray}
& &\langle {\cal A} \{f_{KLM}({\bf u},{\bf x},A){\chi}_{SM_S}{\cal X}_{TM_T}\}
\vert {\cal A'} \{f_{K'L'M'}({\bf u}',{\bf x},A')
{\chi}_{S'M'_S}{\cal X}_{T'M'_T}\}\rangle
\nonumber \\
&=&\delta_{LL'}\delta_{MM'}
\frac{1}{B_{KL}B'_{K'L'}}\left(\frac
{({\rm det}\Gamma)^3}
{(4\pi)^{(N-1)}{\rm det}(
\Gamma-A){\rm det}(\Gamma-A')}
\right)^{3/2} \\
&\times& \left(\sum_{i=1}^{n_o}C_i^{(o)}
\left({(2\pi)^{2N-2} \over {\rm det}(B_i^{(o)}+Q)}\right)^{3/2}
\kappa!\kappa'!
\sum_{k}
{{p_i^{\kappa-k\over 2}{{p_i}'}^{\kappa'-k\over 2}}{q_i}^k \over
({\kappa-k \over 2})! ({\kappa'-k\over 2})! k!} B_{kL} \right), \nonumber
\end{eqnarray}
where the constant term previously suppressed is also included. The
orthogonality of the matrix element in the spin and isospin quantum numbers
is implicitly contained in the coefficients $C_i^{(o)}$.
\par\indent
The calculation of the matrix elements in the way described above is very
simple. This fact becomes especially evident if one compares the work
described above
to the formidable task of the calculation of the matrix elements
in the case where the function $\theta_{LM_L}({\bf x})$
is decomposed into partial waves of the relative coordinates as in Eq. (5).
In fact, in that latter
case one has to integrate over the angles of the relative
coordinates and one has to cope with complicated angular momentum algebra.
We note, however, that the calculation of the matrix element of the latter
type poses no problem if the function $\theta_{LM_L}({\bf x})$ of Eq. (5)
is expressed as a linear combination of the terms of Eq. (6) with
appropriate ${\bf u}$-vectors.
\par\indent
All  the matrix elements can be given in a similar closed analytic
form and the numerical evaluation of the matrix elements as a function
of the nonlinear parameters is therefore straightforward.
The values $\kappa=2K\!\!+\!\!L$ and $\kappa'=2K'\!\!+\!\!L'$ are usually
small in practical cases and the summation over $k$ is limited to just a
few terms in Eq. (99).

\bigskip
\par\indent
\par\indent
This work was supported by OTKA grant No. T17298 (Hungary) and
by a Grant-in-Aid for Scientific Research (No. 05243102 and No. 06640381)
of the Ministry of Education, Science and Culture (Japan). K. V. gratefully
acknowledges the hospitality of the RIKEN LINAC Laboratory and the
support of the Science and Technology Agency of Japan, 1994. The authors
are grateful for the use of RIKEN's VPP500 computer which
made possible most of the calculations.
Thanks are also  due to the support of both Japan Society for the Promotion
of Science and Hungarian Academy of Sciences, 1994-1995. The authors thank
Prof. R. G. Lovas for careful reading of the manuscript and useful
suggestions.

\begin{table}
\caption{List of the parameters of the nucleon-nucleon potentials used in
this paper. The potential consists of a few terms; each is expressed as
$V_if({\mu}_i,r)(W_i+B_iP_{\sigma}-H_iP_{\tau}+M_iP_r)$,
where $P_{\sigma}$, $P_{\tau}$ and $P_r$ are the spin-, isospin- and
space-exchange
operators. The form factor $f({\mu}_i,r)$ is exp($-\mu r^2$) for the
Gaussian potential or exp$(-\mu r)/r$ for the Yukawa potential. The
potential
strength $V$ is in units of MeV and the range $\mu$ in units of fm$^{-2}$
for Gaussian or fm$^{-1}$ for Yukawa, respectively. The Majorana mixture
$M$ of the Volokov potential is set to zero in the calculation. The
Minnesota potential contains the parameter $u$, which is set to unity in
the calculation.}
\begin{tabular}{|c|c|c|c|c|c|c|c|c|}
\hline
Potential & Type & $i$ & $V_i$ & $\mu_i$ & $W_i$ & $M_i$ & $B_i$ & $H_i$ \\
\hline
MT-V   &Yukawa  & 1  & 1458.05  & 3.11 & 1.0  & 0.0  & 0.0 & 0.0       \\
\cite{mt}    &  & 2  & -578.09  & 1.55 & 1.0  & 0.0  & 0.0 & 0.0       \\
\hline
Volkov   & Gauss & 1    & 144.86   &  0.82$^{-2}$ & $1.0-M$  & $M$  &
 0.0     &  0.0      \\
 \cite{Volkov}  & & 2    & $-$83.34   & 1.60$^{-2}$ & $1.0-M$ & $M$ &
 0.0      & 0.0  \\
\hline
ATS3  & Gauss   & 1 & 1000.0   & 3.0  & 1.0  & 0.0 & 0.0 & 0.0       \\
\cite{ATS3} &  & 2 & $-$326.7   & 1.05 & 0.5  & 0.0 & 0.5 & 0.0       \\
        &  & 3    & $-166.0$ & 0.80  & 0.5  & 0.0  & $-0.5$ &  0.0      \\
           &  & 4    & $-43.0$  & 0.60  & 0.5  & 0.0  & 0.5  & 0.0       \\
           &  & 5    & $-23.0$  & 0.40 & 0.5  & 0.0  & $-0.5$ & 0.0       \\
\hline
Minnesota & Gauss  & 1& 200.0 &1.487 &$0.5u$&$1.0-0.5u$&0.0& 0.0       \\
\cite{minesota}&  &2& $-178.0$& 0.639 & $0.25u$& $0.5-0.25u$ & $0.25u$ &
$0.5-0.25u$    \\
         &   & 3 & $-91.85$& 0.465& $0.25u$& $0.5-0.25u$ &$-0.25u$ &
$-0.5+0.25u$
\\ \hline
\end{tabular}
\end{table}

\begin{table}
\caption{Spin (isospin) couplings of the nucleons. The possible
intermediate spin (isospin) configurations of an $N$-nucleon system are
listed above the $(N\!\!-\!\!1)$th horizontal line. The spins
of a 5-nucleon system, for example, are given by the first ten elements of
the  5th column. The spins of the $2-3-4$- particle subsystems of the
5-nucleon system is given in the first ten elements of the $1-2-3$ columns.}
\begin{tabular}{|c|c|c|c|c|c|c|}
\hline
$N$&  2   &  3   &   4   &   5  &  6 & 7    \\  \hline
 & 0   & 1/2  &   0   &  1/2 &  0  &1/2  \\
 & 1   & 1/2  &   0   &  1/2 &  0  &1/2  \\  \cline{0-1}
 & 1   & 3/2  &   1   &  1/2 &  0  &1/2  \\  \cline{0-2}
 & 0   & 1/2  &   1   &  1/2 &  0  &1/2  \\
 & 1   & 1/2  &   1   &  1/2 &  0  &1/2  \\
 & 1   & 3/2  &   2   &  3/2 &  1  &1/2  \\  \cline{0-3}
 & 1   & 3/2  &   1   &  3/2 &  1  &1/2  \\
 & 0   & 1/2  &   1   &  3/2 &  1  &1/2  \\
 & 1   & 1/2  &   1   &  3/2 &  1  &1/2  \\
 & 1   & 3/2  &   2   &  5/2 &  2  &3/2  \\  \cline{0-4}
 & 0   & 1/2  &   0   &  1/2 &  1  &1/2  \\
 & 1   & 1/2  &   0   &  1/2 &  1  &1/2  \\
 & 1   & 3/2  &   1   &  1/2 &  1  &1/2  \\
 & 0   & 1/2  &   1   &  1/2 &  1  &1/2  \\
 & 1   & 1/2  &   1   &  1/2 &  1  &1/2  \\
 & 1   & 3/2  &   2   &  3/2 &  2  &3/2  \\
 & 1   & 3/2  &   1   &  3/2 &  2  &3/2  \\
 & 0   & 1/2  &   1   &  3/2 &  2  &3/2  \\
 & 1   & 1/2  &   1   &  3/2 &  2  &3/2  \\
 & 1   & 3/2  &   2   &  5/2 &  3  &5/2  \\  \cline{0-5}
 & 1   & 3/2  &   2   &  3/2 &  1  &3/2  \\
 & 1   & 3/2  &   1   &  3/2 &  1  &3/2  \\
 & 0   & 1/2  &   1   &  3/2 &  1  &3/2  \\
 & 1   & 1/2  &   1   &  3/2 &  1  &3/2  \\
 & 1   & 3/2  &   2   &  5/2 &  2  &5/3  \\
 & 0   & 1/2  &   0   &  1/2 &  1  &3/2  \\
 & 1   & 1/2  &   0   &  1/2 &  1  &3/2  \\
 & 1   & 3/2  &   1   &  1/2 &  1  &3/2  \\
 & 0   & 1/2  &   1   &  1/2 &  1  &3/2  \\
 & 1   & 1/2  &   1   &  1/2 &  1  &3/2  \\
 & 1   & 3/2  &   2   &  3/2 &  2  &5/2  \\
 & 1   & 3/2  &   1   &  3/2 &  2  &5/2  \\
 & 0   & 1/2  &   1   &  3/2 &  2  &5/2  \\
 & 1   & 1/2  &   1   &  3/2 &  2  &5/2  \\
 & 1   & 3/2  &   2   &  5/2 &  3  &7/2  \\ \hline
\end{tabular}
\end{table}

\begin{table}
\caption{Energies and rms radii
of $N$-nucleon systems interacting via the Malfliet-Tjon
potential V \protect\cite{mt}. The value of ${\cal K}$ in Tables III$-$IX
denotes the basis
dimension beyond which the energies and the radii of the SVM calculation
do not change in the digits shown. ($\hbar^2/m=41.47\, $MeV fm$^2$
throughout Tables III$-$VI).}

\begin{tabular}{lcld@{}ldr}
$N$ & $(L,S)J^{\pi}$&Method  & $E$ (MeV) && $\langle r^2\rangle^{1/2}$ (fm) &
${\cal K}$ \\
\hline
2 &$(0,1)1^{+}$
& Numerical                  & $-$0.4107  &  &3.743    & \\
     && SVM                        & $-$0.4107  &  &3.743    & 5  \\
\hline
3 &$(0,1/2)1/2^{+}$
      & Faddeev\protect\cite{payne,faddeev}   & $-$8.25273   &&   &    \\
      && ATMS\protect\cite{atms}       & $-$8.26&$\pm 0.01$& 1.682  &  \\
      && CHH\protect\cite{hypers}       & $-$8.240&&   &  \\
      && GFMC\protect\cite{gfma}      & $-$8.26&$\pm 0.01$& 1.682 &  \\
      && VMC\protect\cite{wiringa}$^a$ & $-$8.2689 &$\pm 0.03$& 1.68 &  \\
      && SVM                           & $-$8.2527 && 1.682  & 80 \\
\hline
4 &$(0,0)0^{+}$
       & FY\protect\cite{glocka}    & $-$31.36 &&       &     \\
       && ATMS\protect\cite{atms}      & $-$31.36  && 1.40         \\
       && CRCG\protect\cite{crcbgv}  & $-$31.357 &&       & 1000 \\
       && GFMC\protect\cite{gfma}     & $-$31.3&$\pm 0.2$&1.36 &    \\
       && VMC\protect\cite{wiringa}$^a$& $-$31.3 &$\pm 0.05$&1.39 &    \\
       && SVM                          & $-$31.360 && 1.4087  &  150 \\
\hline
5 &$(1,1/2)3/2^{-}$
 & VMC\protect\cite{wiringa}\tablenote{Calculated with Coulomb potential,
the Coulomb contribution then subtracted perturbatively. The potential
strength used in the VMC \cite{wiringa} calculation is $V_1$=1458.25 and
$V_2=-$578.17 MeV, which is slightly different
from that used in the present calculation.}
                                      & $-$42.98 &$\pm$0.16 & 1.51 &  \\
      && SVM                           & $-$43.48 && 1.51 & 500  \\
\hline
6 ($^6$He) & $(0,0)0^{+}$
 & VMC\protect\cite{wiringa}$^a$ & $-$66.34&$\pm$0.29 & 1.50 &  \\
           && SVM                         & $-$66.30 && 1.52   & 800   \\
\hline
7 ($^7$Li)& $(1,1/2)3/2^{-}$
 & SVM                         & $-$83.4 && 1.68    &1300   \\
\end{tabular}
\end{table}

\begin{table}
\caption{Energies and rms radii
of $N$-nucleon systems interacting via the Volkov potential
\protect\cite{Volkov}. The Majorana exchange parameter $M$ is set to zero.}

\begin{tabular}{lcld@{}ldr}
$N$ & $(L,S)J^{\pi}$ & Method   & $E$ (MeV) && $\langle r^2\rangle^{1/2}$ (fm)
& ${\cal K}$\\
\hline
2&$(0,1)1^{+}$   & Numerical                     & $-$0.545  &  &3.44    & \\
      && SVM                           & $-$0.545  &  &3.44    & 5  \\
\hline
3&$(0,1/2)1/2^{+}$  & Faddeev\protect\cite{glocka}& $-$8.43   &&   &  \\
      && Variational\protect\cite{etbm}& $-$8.460  &&1.725   &  \\
      && HH\protect\cite{hh1}          & $-$8.4647  &&   &  \\
      && SVM                           & $-$8.46 && 1.73 & 30  \\
\hline
4&$(0,0)0^+$  & FY\protect\cite{glocka}  & $-$30.27 &&   &     \\
      & & Variational\protect\cite{etbm} & $-$29.490 &&1.47 &     \\
      & & HH\protect\cite{ballot}     & $-$30.3988 &&  &     \\
      && SVM                          & $-$30.42 && 1.49  &  50 \\
\hline
5&$(1,1/2)3/2^-$&SVM                           & $-$43.00 && 1.59 & 120  \\
\hline
6 ($^6$Li)&$(0,1)1^{+}$  & SVM              & $-$66.25 && 1.60   & 250   \\
\hline
7 ($^7$Li)&$(1,1/2)3/2^{-}$  & SVM         & $-$98.75 && 1.57    & 400   \\
\end{tabular}
\end{table}

\begin{table}
\caption{Energies and rms radii
of $N$-nucleon systems interacting via the Afnan-Tang S3 potential
\protect\cite{ATS3}.}

\begin{tabular}{lcld@{}ldr}
$N$&$(L,S)J^{\pi}$ & Method   & $E$ (MeV) && $\langle r^2\rangle^{1/2}$ (fm) &
${\cal K}$\\
\hline
2&$(0,1)1^+$    & Numerical                     & $-$2.216  &  &1.94    & \\
      && SVM                           & $-$2.216  &  &1.94    & 7  \\
\hline
3&$(0,1/2)1/2^+$  & Faddeev\protect\cite{glocka}    & $-$8.20  &&    &  \\
      && Faddeev\protect\cite{friar}   & $-$8.765   &&   &  \\
      && GFMC\protect\cite{gfmb}    & $-$8.73 &$\pm 0.10$& 1.72  &  \\
      && Variational\protect\cite{etbm} & $-$6.677 && 1.727  &  \\
      && SVM                           & $-$8.753 && 1.67  & 40  \\
\hline
4&$(0,0)0^{+}$    & FY\protect\cite{glocka}   & $-$28.80 &&  &  \\
      & &Variational\protect\cite{etbm}    & $-$25.654 &&1.44  &  \\
         & & SVM                        & $-$30.37 && 1.42  & 140 \\
\hline
5&$(1,1/2)3/2^{-}$    & SVM              & $-$44.27 && 1.48  & 220  \\
\hline
6($^6$Li)&$(0,1)1^+$   & SVM             & $-$70.65 && 1.49  & 400   \\
\end{tabular}
\end{table}

\begin{table}
\caption{Energies and rms radii of $N$-nucleon systems interacting via
 the Minnesota potential \protect\cite{minesota} with the exchange
parameter $u=1$. The Coulomb interaction is included ($e^2=1.44$ MeV fm).
The experimental
$\langle r^2\rangle^{1/2}$ value is the point charge radius with the
proton's finite size corrected.}

\begin{tabular}{lclddr}
$N$&$(L,S)J^{\pi}$ & Method                  & $E$ (MeV) & $\langle
r^2\rangle^{1/2}$
(fm) & ${\cal K}$\\
\hline
2&$(0,1)1^+$     & SVM                   & $-$2.202    &1.952    & 5  \\
       && Exp.            & $-$2.224    &1.96    &   \\
\hline
3 ($^3$H)&$(0,1/2)1/2^+$    & SVM                     & $-$8.380   & 1.698  &
40 \\
       && Exp.                           & $-$8.481    & 1.57  &   \\
\hline
4 ($^4$He) &$(0,0)0^+$ & SVM              & $-$29.937 & 1.41     &  60 \\
            && Exp.       & $-$28.295       & 1.47        &     \\
\hline
5 &$(1,1/2)3/2^-$    & SVM                          & unbound &      &  \\
       && Exp.                         & unbound &      &  \\
\hline
6 ($^6$He) &$(0,0)0^+$ & SVM                         & $-$30.07 & 2.44  & 600
\\
            && Exp.                  & $-$29.271 &    &     \\
\ \  ($^6$Li) &$(0,1)1^+$ & SVM                         & $-$34.59 & 2.22  &
600   \\
            && Exp.                  & $-$31.995 & 2.43   &     \\
\end{tabular}
\end{table}

\begin{table}
\caption{Energies and rms radii of $N$-$\alpha$ systems interacting via
 the Ali-Bodmer potential \protect\cite{AB} of Eq. (63). The
$\alpha$-particle is
considered to be structureless boson. ($\hbar^2/M_{\alpha}=41.467/4\, $MeV
fm$^2$).}

\begin{tabular}{llddr}
$N$ & Method                  & $E$ (MeV) & $\langle r^2\rangle^{1/2}$
(fm) & ${\cal K}$\\
\hline
2     &                           & unbound   &    &           \\
\hline
3      & ATMS\cite{atms}          & $-$5.18     & 2.43  &      \\
       & SVM                      & $-$5.18     & 2.43 & 60   \\
\hline
4      & ATMS\cite{atms}          & $-$11.1     & 2.65     &   \\
       & SVM                      & $-$11.07    & 2.65    &  150   \\
\hline
5      & SVM                      & $-$16.22    & 2.99     & 400 \\
\hline
6      & SVM                      & $-$20.13    & 3.32     & 600   \\
\end{tabular}
\end{table}

\begin{table}
\caption{Energies and rms
radii of electron-positron systems treated as fermions (f) and
as bosons (b). Atomic units are used.}
\begin{tabular}{llddr}
System & Method                  & $E$  & $\langle r^2\rangle^{1/2}$ & ${\cal
K}$\\
\hline
($e^+,e^-$) b,f   & SVM                   & $-$0.25  &1.732    & 10  \\
          & Exact        & $-$0.25    &1.732    &     \\
\hline
($2e^+,e^-$) b,f   & SVM                  & $-$0.262004   & 4.592       &
150 \\
            & Variational\protect\cite{epe}   & $-$0.2620050702325  & 4.594  &
700  \\
         & Faddeev \protect\cite{faddeev} & $-$0.26202  &  &  \\
\hline
($2e^+,2e^-$) b,f  & SVM           &$-$0.515989  & 3.608 &  300 \\
            & Variational\protect\cite{epep,KA93}  & $-$0.515980  & 3.600 &
300 \\
\hline
($3e^+,2e^-$) f   & SVM                         & unbound  &  & 1000  \\
($3e^+,2e^-$) b   & SVM                         & $-$0.5493 & 3.53  & 200  \\
\hline
($3e^+,3e^-$)  f & SVM                         & unbound &  & 1000     \\
($3e^+,3e^-$)  b & SVM                         & $-$0.820 & 3.42 & 300   \\
                & Variational\protect\cite{boson}       & $-$0.789 &      &   5
  \\
\end{tabular}
\end{table}

\begin{table}
\caption{Energies and rms
radii of ``self-gravitating" $m$-particle--$n$-antiparticle systems
($m+,n-$); f: fermions; b: bosons.
VLB and VUB stand for the variational lower and upper bounds given in
Ref. \protect\cite{selfgrav}. The units of the energy and length
are $G^2 m^5\hbar^{-2}$ and  $G^{-1} m^{-3} \hbar^2$, respectively.}
\begin{tabular}{llddlr}
System           & Method      & $E$      &           &
\hspace{-5pt}$\langle r^2\rangle^{1/2}$
  & ${\cal K}$\\
\hline
($+,-$) b,f  & SVM         & $-$0.25  &\hspace{-8pt}
&\hspace{-5pt}1.732    & 10  \\
                 & Exact       & $-$0.25  &\hspace{-8pt}
 &\hspace{-5pt}1.732    &     \\
\hline
($2+,-$) b,f & SVM         & $-$1.072 &\hspace{-8pt}
&\hspace{-5pt}1.304& 15 \\
                 &
       Variational\protect\cite{selfgrav} & $-$1.067 &\hspace{-8pt}           &
     &   \\
\hline
($2+,2-$) b,f& SVM         & $-$2.791 &\hspace{-8pt}
&\hspace{-5pt}1.027&  100 \\
                 & VUB \ (VLB) & $-$1.951 &\hspace{-8pt} ($-$3.00) &      &  \\
\hline
($3+,2-$) f  & SVM         & $-$3.758 &\hspace{-8pt}
&\hspace{-5pt}1.554& 200  \\
($3+,2-$) b  & SVM         & $-$5.732 &\hspace{-8pt}
&\hspace{-5pt}0.844& 200  \\
                 & VUB \ (VLB) & $-$4.336 &\hspace{-8pt} ($-$6.25) &      &  \\
\hline
($3+,3-$) f  & SVM         & $-$6.409 &\hspace{-8pt}
&\hspace{-5pt}1.621& 300   \\
($3+,3-$) b  & SVM         & $-$10.215&\hspace{-8pt}
&\hspace{-5pt}0.718& 300 \\
                 & VUB \ (VLB) & $-$8.130 &\hspace{-8pt} ($-$11.25)&      &
 \\
\end{tabular}
\end{table}
\newpage
\noindent
Table X. Components of the two-particle interaction matrix elements.
The operators $B^p$ and $C^{p}_{lm}$ are defined in Eq. (70) of
Appendix B. The symbol $p$ specifies the component of the two-body
interaction. The following additional notations are
introduced: ${\bf r}_{12}={\bf r}_1-{\bf r}_2$,
${\bf p}_{12}={1\over 2\hbar}({\bf p}_1-{\bf p}_2)$,
${\bf L}={\bf r}_{12}\times{\bf p}_{12}$,
${\bf S}={1 \over 2}$ ({\mbox{\boldmath $\sigma$}}$_1$+
{\mbox{\boldmath $\sigma$}}$_2$),
$S_{12}=3$({\mbox{\boldmath $\sigma$}}$_1\cdot
{\hat {\bf r}_{12}})$
({\mbox{\boldmath $\sigma$}}$_2 \cdot
{\hat {\bf r}_{12}})-$
{\mbox{\boldmath $\sigma$}}$_1 \cdot$
{\mbox{\boldmath $\sigma$}}$_2$,
${\bf x}={{\bf s}'}_1-{{\bf s}'}_2$,
and
$({\bf x})_m=\sqrt{4\pi\over 3} x Y_{1m}({\hat{\bf x}})$.
\squeezetable
\begin{table}
\begin{tabular}{|c|c|c|c|c|c|}\hline
& & & & & \\
$p$ & definition
&$ B^p$ &$ C^{p}_{00}$     &$ C^{p}_{1m} $&$ C^{p}_{2m}$ \\
& & & & & \\ \hline
$c$ & 1          & 0 &$ \sqrt{4\pi}$& 0     & 0      \\
& & & & & \\ \hline
& & & & & \\
$t$ & $S_{12}$ &
0 & 0 & 0 &
$\sqrt{24 \pi \over 5} [${\mbox{\boldmath $\sigma$}}$_1\times$
{\mbox{\boldmath $\sigma$}}$_2]^{(2)}_m$\\
& & & & & \\ \hline
& & & & & \\
b & ${\bf L} \cdot {\bf S}$& 0 & 0 &
$-\sqrt{{2\pi\over 3}}\nu\Bigl[{\bf x}\times$
(
{\mbox{\boldmath $\sigma$}}$_1$+
{\mbox{\boldmath $\sigma$}}$_2)\Bigr]^{(1)}_{m}$ & 0 \\ & & & & &
\\ \hline
& & & & & \\
$q$   & ${\bf L}^2$   &
$-{2\over 3} \nu^2 {\bf x}^2$ & 0  &
$ \sqrt{16 \pi \over 3} \nu ({\bf x})_{m}$ &
$ \sqrt{8 \pi \over 15} \nu^2 [{\bf x}\times
{\bf x}]^{(2)}_{m}$ \\
& & & & & \\
\hline
& & & & & \\
$q$  & ${\bf p}_{12}^2 $ & $\nu^2$ &
$\sqrt{4\pi} \nu(-3 +\nu{\bf x}^2)$ &
$-\sqrt{16\pi \over 3}\nu^2 ({\bf x})_{m}$
& 0 \\
& & & & & \\
\hline
& & & & & \\
$bb$ & $({\bf L} \cdot {\bf S})^2$ &
$-{1\over 6} \nu^2\Bigl\lbrace{\bf x}^2 (2+$
{\mbox{\boldmath $\sigma$}}$_1\cdot$
{\mbox{\boldmath $\sigma$}}$_2)-$ & 0 &
$ \sqrt{\pi\over 3}\nu \Bigl\lbrace({\bf x})_{m}(2+
${\mbox{\boldmath $\sigma$}}$_1\cdot$
{\mbox{\boldmath $\sigma$}}$_2) +$ &
$\sqrt{2\pi\over 15}\nu^2\Bigl\lbrace[{\bf x}\times
{\bf x}]^{(2)}_{m}+$ \\
 & & $({\bf x} \cdot
${\mbox{\boldmath $\sigma$}}$_1)$
$({\bf x} \cdot $
{\mbox{\boldmath $\sigma$}}$_2) \Bigr\rbrace$& &
${1\over \sqrt{2}}
\Bigl[{\bf x}\times
(${\mbox{\boldmath $\sigma$}}$_1+$
{\mbox{\boldmath $\sigma$}}$_2)\Bigr]^{(1)}_{m}$ &
$2 \Bigl[[{\bf x}\times
${\mbox{\boldmath $\sigma$}}$_1]^{(1)}\times
[{\bf x}\times
${\mbox{\boldmath $\sigma$}}$_2]^{(1)}\Bigr]^{(2)}_{m}\Bigr\rbrace$
\\
 & & & &
$-{1\over 2}(
${\mbox{\boldmath $\sigma$}}$_{1m}({\bf x}\cdot
${\mbox{\boldmath $\sigma$}}$_2)+
${\mbox{\boldmath $\sigma$}}$_{2m}({\bf x}\cdot
${\mbox{\boldmath $\sigma$}}$_1))\Bigr\rbrace$
& \\  & & & & & \\ \hline \hline
\end{tabular}
\end{table}
\newpage
\begin{center}
{\large  Figure Captions}
\end{center}
\bigskip
\begin{enumerate}
\item[ Fig. 1]: Different sets of relative coordinates for a system of six
identical particles.
\item[Fig. 2]:  Convergence of the $^6$Li energy on different
random paths.  The Volkov potential \cite{Volkov} is used.
\item[Fig. 3]:  Convergence of the $\alpha$-particle energy for
the potentials listed in  Table I.
\end{enumerate}


\begin{thebibliography}{99}
\bibitem{payne} J. L. Friar, B. F. Gibson, and G. L. Payne, Phys. Rev. C{\bf
24}, 2279 (1981).
\bibitem{faddeev} N. W. Schellingerhout and L. P. Kok, Nucl. Phys.
{\bf A508}, 299c (1990).
\bibitem{glockb} W. Gl\"ockle and H. Kamada, Phys. Rev. Lett. {\bf 71},
971 (1993).
\bibitem{atms} Y. Akaishi, in {\it International Review of Nuclear Physics},
Vol. {\bf 4}, (World Scientific, Singapore, 1986), p.259.
\bibitem{hypers} A. Kievsky, M. Viviani, and S. Rosati, Nucl. Phys.
{\bf A501}, 503 (1989); ibid.
{\bf A551}, 241 (1993); ibid. {\bf A577}, 511 (1994).
\bibitem{crcbgv} M. Kamimura and H. Kameyama, Nucl. Phys.  {\bf A508} 17c,
(1990); H. Kameyama, M. Kamimura, and Y. Fukushima, Phys. Rev. C{\bf40},
974, (1989).
\bibitem{mcv} J. Carlson and V. R. Pandharipande, Nucl. Phys. {\bf A371},
301 (1981).
\bibitem{vmc} R. B. Wiringa, Nucl. Phys. {\bf A543}, 199c (1992).
\bibitem{gfma} J. G. Za\-bo\-litzky and M. H. Kalos, Nucl. Phys. {\bf A356},
114 (1981);
 J. G. Za\-bo\-litzky, K. E. Schmidt, and M. H. Kalos, Phys. Rev.
 C{\bf 25}, 1111 (1982).
\bibitem{gfmb} J. Carlson, Phys. Rev. C{\bf 36}, 2026 (1987).
\bibitem{prl95}B. S. Pudliner, V. R. Pandharipande, J. Carlson, and R. B.
Wiringa, Phys. Rev. Lett. {\bf 74}, 4396 (1995).
\bibitem{SVM} V. I. Kukulin and V. M. Krasnopol'sky, J. Phys. G{\bf 3}, 795
(1977).
\bibitem{VSL} K. Varga, Y. Suzuki, and R. G. Lovas, Nucl. Phys. {\bf A371},
447 (1994).
\bibitem{corrgauss} S. F. Boys, Proc. R. Soc. London Ser. A{\bf 258},
402 (1960);
 K. Singer, ibid. A{\bf 258}, 412 (1960).
\bibitem{temper} S. A. Alexander, H. J. Monkhorst, and K. Szalewicz,
J. Chem. Phys. {\bf 85}, 5821 (1986).
\bibitem{etbm} S. Fantoni, L. Panattoni, and S. Rosati, Nouvo Cimento
A{\bf 69}, 81 (1970).
\bibitem{SVAO} K. Varga, Y. Suzuki, and Y. Ohbayasi, Phys. Rev. C{\bf
50}, 189 (1994); K. Arai, Y. Suzuki, and K. Varga, ibid. C{\bf
51}, 2488 (1995); K. Varga, Y. Suzuki, and I. Tanihata, submitted to
Phys. Rev. C.
\bibitem{KA91}  P. M. Kozlowski and L. Adamowicz, J. Chem. Phys. {\bf 95}, 6681
(1991).
\bibitem{epep} D. B. Kinghorn and R. D. Poshusta, Phys. Rev.
A{\bf 47}, 3671 (1993).
\bibitem{SD} H. De Raedt and M. Frick, Phys. Rep. {\bf 231}, 109   (1993).
\bibitem{suzuki} Y. Suzuki, E. J. Reske, and K. T. Hecht, Nucl. Phys. {\bf
A381}, 77 (1982).
\bibitem{Brink} D. Brink, {\it Proceedings of the International School of
Physics ``Enrico Fermi''}, Course {\bf 36} (1965), ed. C. Bloch,
(Academic Press, New York and London, 1966), p.247.
\bibitem{wild} K. Wildermuth and Y. C. Tang, {\it A Unified Theory of the
Nucleus, Clustering Phenomena in Nuclei}, Vol. 1, eds. K. Wildermuth and
P. Kramer (Vieweg, Braunschweig, Germany, 1977).
\bibitem{supple} H. Horiuchi, Prog. Theor. Phys. Suppl. No. {\bf 62}, 90
(1977).
\bibitem{feldmeier} H. Feldmeier, Nucl. Phys. {\bf A515}, 147 (1990).
\bibitem{horiuchi}A. Ono, H. Horiuchi, T. Maruyama, and A. Ohnishi, Phys. Rev.
C{\bf 47}, 2652 (1993).
\bibitem{Lowden} P. O. L\"owdin, Phys. Rev. {\bf 97}, 1490 (1955).
\bibitem{Kami} M. Kamimura, Prog. Theor. Phys. Suppl. No. {\bf 62},
236 (1977).
\bibitem{mt} R. A. Malfliet and J. A. Tjon, Nucl. Phys. {\bf A127},
161 (1969).
\bibitem{Volkov} A. B. Volkov, Nucl. Phys. {\bf 74}, 33 (1965).
\bibitem{ATS3} I. R. Afnan and Y. C. Tang, Phys. Rev. {\bf 175}, 1337 (1968).
\bibitem{minesota} D. R. Thompson, M. LeMere, and Y. C. Tang,
Nucl. Phys. {\bf A286}, 53 (1977); I. Reichstein and Y. C. Tang, Nucl. Phys.
{\bf A158}, 529 (1970).
\bibitem{glocka} H. Kamada and W. Gl\"ockle, Nucl. Phys. {\bf A548}, 205
(1992).
\bibitem{wiringa} R. B. Wiringa, private communication.
\bibitem{hh1} G. Erens, J. L. Visschers, and R. van Wageningen, Ann. Phys.
(N. Y.) {\bf 67}, 461 (1971).
\bibitem{ballot} J. L. Ballot, Z. Phys. A$-$Atoms and Nuclei {\bf 302}, 347
(1981).
\bibitem{friar} J. L. Friar, quoted in \cite{gfmb}.
\bibitem{LKBD} R. G. Lovas, A. T. Kruppa, R. Beck, and F. Dickmann, Nucl. Phys.
{\bf A474} 451, (1987).
\bibitem{AB} S. Ali and A. R. Bodmer, Nucl. Phys. {\bf 80}, 99 (1966).
\bibitem{wheeler} J. A. Wheeler, Ann. N. Y. Acad. Sci. {\bf 48}, 219 (1946).
\bibitem{epe} A. Yeremin, A. M. Frolov, and E. B. Kutukova, Few-Body Systems
{\bf 4}, 111 (1988).
\bibitem{ho}  Y. K. Ho, J. Phys. B{\bf 16}, 1503 (1983).
\bibitem{mills} A. P. Mills, Phys. Rev. Lett. {\bf 46}, 717 (1981).
\bibitem{hyll} E. A. Hylleraas and A. Ore, Phys. Rev. {\bf 71}, 493 (1947);
A. Ore, ibid. {\bf 71}, 913 (1947); ibid. {\bf 83}, 665 (1951).
\bibitem{frolov95} A. M. Frolov, S. I. Kryuchkov, and V. H. Smith, Jr.,
Phys. Rev. A{\bf 51}, 4514 (1995).
\bibitem{KA93}  P. M. Kozlowski and L. Adamowicz, Phys. Rev. A{\bf 47}, 1903
(1993).
\bibitem{boson} S. Fleck and J.-M. Richard, Few-body systems, in press.
\bibitem{ho86} Y. K. Ho, Phys. Rev. A{\bf 33}, 3584 (1986).
\bibitem{selfgrav} J. L. Basdevant, A. Martin, and J. M. Richard,
Nucl. Phys. {\bf B343}, 60 (1990).
\bibitem{dtmu} M. Kamimura, Phys. Rev. A{\bf 38}, 621 (1988).
\bibitem{LP81} I. E. Lagaris and V. R. Pandharipande, Nucl. Phys.
{\bf A539}, 331 (1981).
\bibitem{WSA84} R. B. Wiringa, R. A. Smith, and T. L. Ainsworth, Phys. Rev.
C{\bf 29}, 1207 (1984).
\end{thebibliography}
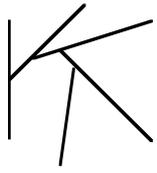
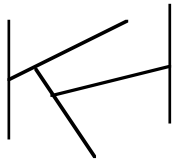
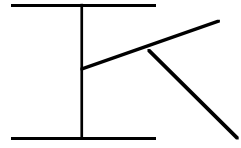
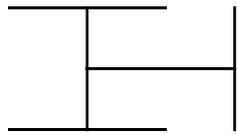
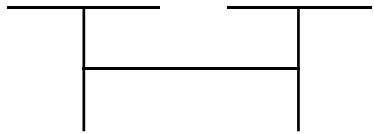
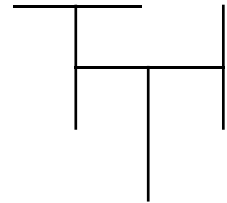
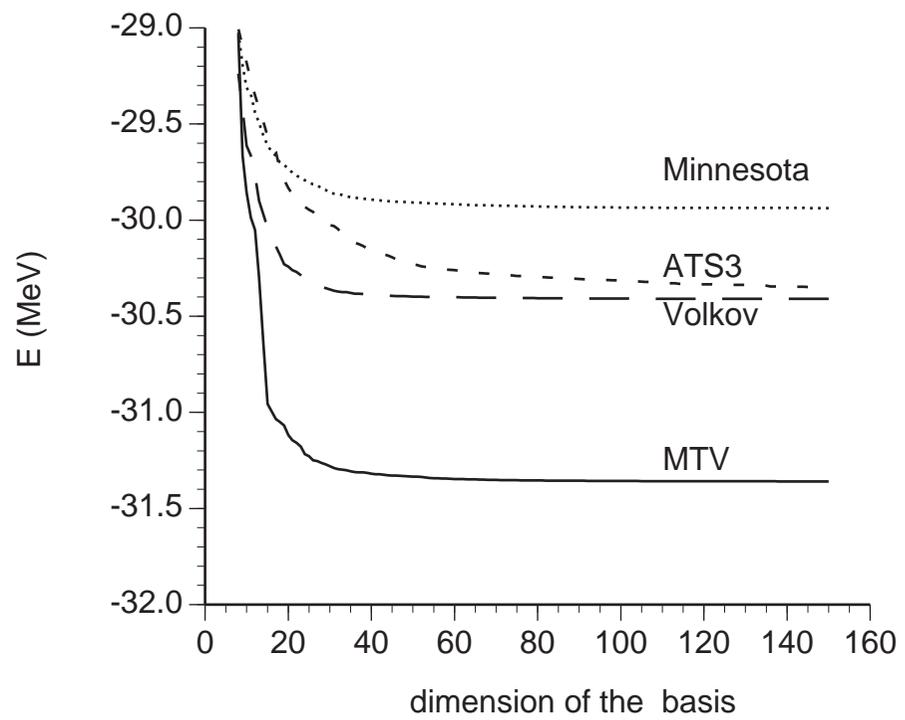
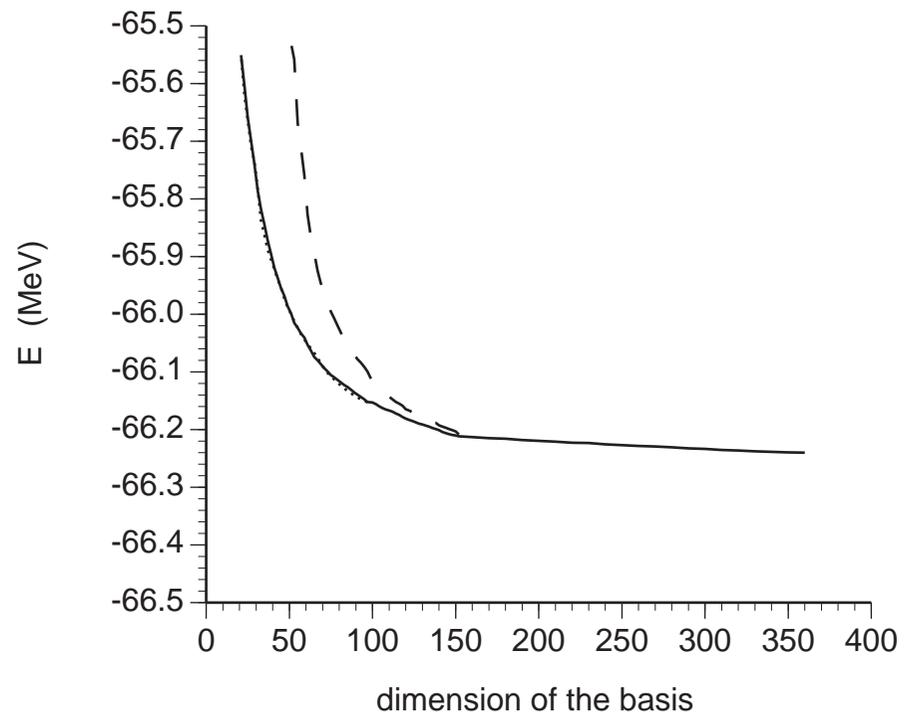
\end{document}